\documentclass[twocolumn]{aastex631}

\submitjournal{ApJ}
\usepackage{graphicx}

\shorttitle{Satellites of Elliptical Galaxies}
\shortauthors{Kim et al.}

\def\kms{\rm\thinspace km~s^{-1}}
\def\Msun{M_{\odot}}
\def\Mstel{M_{\ast}}
\def\Mvir{M_{\rm vir}}
\usepackage{soul}

\begin{document}

\title{Investigating the Impact of Supernova Feedback on Satellites in Elliptical Galaxies}
\author{Sumi Kim}
\affiliation{Department of Physics, University of Seoul, 163 Seoulsiripdaero, Dongdaemun-gu, Seoul 02504, Republic of Korea}

\correspondingauthor{Ena Choi}
\email{enachoi@uos.ac.kr}
\author[0000-0002-8131-6378]{Ena Choi}
\affiliation{Department of Physics, University of Seoul, 163 Seoulsiripdaero, Dongdaemun-gu, Seoul 02504, Republic of Korea}

\author[0000-0001-8481-2660]{Amanda C. N. Quirk}
\affil{Department of Astronomy, Columbia University, 
       550 West 120th Street, 
       New York, NY 10027, USA}

\author[0000-0002-6748-6821]{Rachel S. Somerville}
\affil{Center for Computational Astrophysics, Flatiron Institute, 162 5th Ave., 
        New York, NY 10010, USA}

\author[0000-0002-7314-2558]{Thorsten Naab}
\affil{Max-Planck-Institut f\"ur Astrophysik,
        Karl-Schwarzschild-Strasse 1, 85741 Garching, Germany}
        
\author[0000-0002-6405-9904]{Jeremiah P. Ostriker}
\affil{Department of Astronomy, Columbia University, 
       550 West 120th Street, 
       New York, NY 10027, USA}
\affil{Department of Astrophysical Sciences, Princeton University, 
       Princeton, NJ 08544, USA}
    
\author[0000-0002-3301-3321]{Michaela Hirschmann}
\affil{Institute for Physics, Laboratory for Galaxy Evolution and Spectral modelling, Ecole Polytechnique Federale de Lausanne,\\
Observatoire de Sauverny, Chemin Pegasi 51, 1290 Versoix, Switzerland}
\affil{INAF - Osservatorio Astronomico di Trieste, via G. B. Tiepolo 11, I-34143 Trieste, Italy}

\begin{abstract}
We investigate the influence of supernova (SN) feedback on the satellites of elliptical host galaxies using hydrodynamic simulations. Utilizing a modified version of the GADGET-3 code, we perform cosmological zoom-in simulations of 11 elliptical galaxies with stellar masses in the range $10^{11} \Msun < \Mstel < 2 \times 10^{11} \Msun$. We conduct two sets of simulations with identical initial conditions: the Fiducial model, which includes a three-phase SN mechanical wind, and the weak SN feedback model, where nearly all SN energy is released as thermal energy with a reduced SN wind velocity. Our comparison shows minimal differences in the elliptical host galaxies, but significant variations in the physical properties of satellite galaxies. The weak SN feedback model produces a larger number of satellite galaxies compared to the Fiducial model, and significantly more than observed. 
For satellite galaxies with stellar masses above  $10^{8}$ $\Msun$, the weak SN feedback model generates approximately five times more satellites than observed in the xSAGA survey. Most of these overproduced satellites have small stellar masses, below $10^{10}$ $\Msun$. Additionally, satellites in the weak SN feedback model are about 3.5 times more compact than those observed in the SAGA survey and the Fiducial model, with metallicities nearly 1 dex higher than observed values. In conclusion, the satellite galaxies in the Fiducial model, which includes mechanical SN feedback, exhibit properties more closely aligned with observations. This underscores the necessity of incorporating both mechanical AGN and SN feedback to reproduce the observed properties of elliptical galaxy and their satellites in simulations.

\end{abstract}

\section{Introduction}\label{sec:intro}
In our modern cosmological paradigm, galaxies are thought to form through a series of mergers as well as by accreting gas. The hierarchical formation history of galaxies and their host dark matter halos lead to gravitationally bound systems of galaxies, in which satellite galaxies orbit a central galaxy.
Specifically, the satellite number count of the Milky Way (MW), with the deepest existing observations of satellite galaxies, has been compared to simulations, revealing the ``missing satellite problem'' \citep{1999ApJ...522...82K,2004ApJ...609..482K,2008MNRAS.391.1685S}. Initially, the number of satellite galaxies in simulated hosts with the same mass as the MW was ten times greater than the number observed in the MW \citep[e.g.][]{1999ApJ...524L..19M}. However, more sensitive observations have since increased the known satellite count of the MW \citep{2015ApJ...807...50B,2015ApJ...805..130K}, while simulations that incorporate baryonic physics have decreased the predicted observable satellite count to a level that is in agreement with observations \citep[e.g.][]{2000ApJ...539..517B,2013ApJ...765...22B,2016ApJ...827L..23W,2014Natur.509..177V,Rohr2024}.

The spatial distribution of satellites around central galaxies is a valuable tool for testing galaxy formation physics \citep[see][for recent reviews]{2017ARA&A..55..343B,Hui2022}. Numerous simulations have demonstrated that baryonic processes, such as stellar winds \citep{2019MNRAS.482.1304E,Costa2019}, cosmic ionization \citep{2008MNRAS.390..920O, 2020MNRAS.494.2200K}, and supernova (SN) winds \citep{2014ApJ...792...99S,2015MNRAS.451.2900K}, significantly suppress star formation in small galaxies therefore can reduce number counts of small satellite galaxies \citep{2012ApJ...761...71Z,2013ApJ...765...22B,2014ApJ...786...87B,2016MNRAS.457.1931S,2012ApJ...752L..19Q}.  These simulations studying satellite galaxies and comparing them to observations primarily focus on MW-analog hosts \citep[e.g.][]{2009MNRAS.399L.174O,2010MNRAS.406..208O,2013MNRAS.429..633G,2020MNRAS.491.1471S,2022MNRAS.511.1544F}, because the MW and M31 are the only large galaxies in the Local Group where such comparative analyses of satellite properties are feasible \citep{2006ApJ...647L.111B,2006MNRAS.365.1263M,2008ApJ...686..279K,2012AJ....144....4M}.

Recently, however, more observational data have become available for comparing the properties of satellite galaxies in more massive host galaxies \citep[e.g.][]{2012ApJ...752...99N,2011MNRAS.412.2498M,2012MNRAS.422.2187M,2013MNRAS.429..792M,2014MNRAS.442..347R}. For example, \citet{2015MNRAS.454.1605R} utilized the spectroscopic catalog of the SDSS to investigate the satellite abundance of local massive galaxies with masses greater than $10^{11} \Msun$. They examined the physical properties of satellites with masses greater than $10^9 \Msun$ located within 300 kpc of their host galaxies. More recently, \citet{Wu2022} employed a convolutional neural network (CNN) to select nearly 11,000 satellites of 7,542 host galaxies with masses $10^{9.5} < M_{*}/\Msun < 10^{11}$ from the DESI Legacy Imaging surveys and analyze their physical properties.

Despite the accumulation of these observations, only a few studies have analyzed the properties of satellites of massive host galaxies with masses greater than the MW, specifically in the $10^{11-12} \Msun$ range in simulations \citep{2012ApJ...752L..19Q}. Massive galaxies in the $10^{11-12} \Msun$ range are believed to have formed through a two-phase process \citep{2007ApJ...658..710N,Oser2010,2013ApJ...768L..28H}, with the outer regions formed via dry minor mergers with nearby satellite galaxies (\citealp{2009ApJ...699L.178N}, \citealp{2010MNRAS.401.1099H}, \citealp{2013MNRAS.429.2924H}, \citealp{2015MNRAS.449..528H}, \citealp{Choi2018}, see \citealp{2017ARA&A..55...59N} for a review). Therefore, it is of particular interest to investigate the properties of satellites of massive host galaxies. The building blocks that form a merged massive galaxy are not easily distinguishable from the parent galaxy itself. However, analyzing the physical properties of satellite galaxies, which serve as analogs for these smaller building blocks, can provide deeper insights into the formation processes of massive galaxies.

In this study, we investigate the impact of SN feedback models on the satellite galaxy properties of massive galaxies. While in-situ star formation and quenching in massive galaxies with deep gravitational potentials are primarily influenced by AGN feedback rather than SN feedback (e.g., \citealp{2000MNRAS.311..576K}, \citealp{2003ApJ...599...38B}, \citealp{2008MNRAS.391..481S}, \citealp{2009MNRAS.398...53B}, \citealp{2016MNRAS.463.3948D}, see \citealp{2015ARA&A..53...51S} for a review), the latter still plays a significant role in smaller halo mass ranges (\citealp{2008MNRAS.389.1137S}, \citealp{2013MNRAS.434.3142A}, \citealp{Booth2013}, cf., see \citealp{Dashyan2019} for AGN feedback effects on small halos). Therefore, SN feedback has a substantial impact on the properties of the smaller building blocks that constitute massive galaxies. Although SN feedback may not strongly affect the physical properties of the massive host galaxy directly, it exerts an indirect influence by shaping these smaller building blocks. By analyzing the effects of SN feedback on the physical properties, distribution, and count of satellites—analogous to these smaller building blocks—we aim to gain a clearer understanding of their role in massive galaxy formation.

In this work, we use cosmological hydrodynamic simulations to study the effect of SN feedback on the physical properties of satellite galaxies of massive host galaxies. We conduct two zoom-in simulation suites with different SN feedback models and compare the physical properties of the simulated satellite galaxies with the observations of \citet{2015MNRAS.454.1605R} and \citet{Wu2022}. The first set of simulations, the Fiducial model, includes a three-phase SN mechanical wind \citep{2017ApJ...836..204N}, while the second set, the weak SN feedback models have reduced the SN wind velocity and release almost all of the SN energy as thermal energy. Mechanical SN feedback models are known to more effectively couple with the surrounding gas, driving galactic winds and more efficiently regulating star formation compared to thermal feedback models \citep{2013ApJ...770...25A,2014MNRAS.445..581H,2015MNRAS.451.2900K,2015ApJ...802...99K,2015MNRAS.451.2757W}. Numerous studies have demonstrated that mechanical SN feedback is crucial for accurately reproducing galaxies with masses similar to or smaller than the Milky Way \citep[e.g.][]{2015ApJ...809...69S,2018MNRAS.477.1578H}. In this paper, we explore the effects of SN feedback on satellite galaxies, and specifically compare how mechanical and thermal SN feedback models influence the physical properties of satellite galaxies of massive halos.

This paper is organized as follows: In Section~\ref{sec:model}, we describe the simulations and explain the criteria for defining and selecting satellite galaxies. In Section~\ref{sec:result}, we compare the physical properties of satellites from two different SN feedback models and discuss how they align with observations.  Section~\ref{sec:discussion} presents the discussion, and concludes the paper with a summary.

\section{Simulation and Methods} \label{sec:model}
\subsection{Galaxy formation simulations}

We utilized two sets of 11 cosmological zoom-in hydrodynamic simulations encompassing a massive halo, each with present-day halo masses ranging $1.4 \times 10^{12} < M_{\mathrm{vir}}/\Msun < 6.6 \times 10^{12}$ and present-day stellar masses ranging $10^{11} < \Mstel/\Msun < 2 \times 10^{11}$.

The simulation employed in our study was first introduced in \citet{Choi2017}. The physical processes included in the simulations encompass star formation, mechanical supernova (SN) feedback, wind feedback from massive stars, feedback from asymptotic giant branch (AGB) stars, as well as metal cooling and diffusion. In addition, simulations incorporate a novel approach of mechanical and radiative active galactic nucleus (AGN) feedback, implemented in a self-consistent manner, resulting in the launch of high-velocity galactic outflows. 

To investigate the impact of SN feedback on satellite formation around elliptical galaxies, we conduct additional cosmological zoom-in simulations using the same initial conditions but with the SN feedback artificially suppressed to approximate the effect of thermal SN feedback.

(1) Fiducial: This set of simulations mirrors the Fiducial model previously presented in \cite{Choi2017}.

(2) Weak SN: 
These simulations are identical to the Fiducial set but with substantially reduced SN feedback. Specifically, the outflow velocity from supernova winds is set to $v_{\rm out,SN}=10 \kms$ in order to approximate the lower efficiency of thermal SN feedback, while the fiducial model adopts $v_{\rm out,SN}=4500 \kms$ for mechanical feedback.

In the following subsections, we provide a brief overview of the simulation characteristics, with a particular focus on SN feedback. For more details, we direct interested readers to \cite{Choi2017}. 

\subsubsection{Code basics and initial conditions}
The simulation is conducted with the code SPHGal \citep{2014MNRAS.443.1173H}, a modified version of the parallel smoothed particle hydrodynamics (SPH) code GADGET-3 \citep{2005MNRAS.364.1105S}. SPHGal adopts a number of improvements to overcome the numerical fluid-mixing problems of classical SPH codes have \citep[e.g.][]{Agertz2007}: a density-independent pressure-entropy SPH formulation \citep{2001MNRAS.323..743R,2013ApJ...768...44S,2013MNRAS.428.2840H}, an artificial thermal conductivity \cite{2012MNRAS.422.3037R}, an improved artificial viscosity implementation \citep{2010MNRAS.408..669C}, and a Wendland $C^4$ kernel with 200 neighboring particles \citep{2012MNRAS.425.1068D}. The code also incorporates a time-step limiter designed to decrease the time-step of adjacent particles in the vicinity of rapidly moving particles \citep{2009ApJ...697L..99S,2012MNRAS.419..465D}. This enhancement ensures a more accurate modeling of shock propagation and feedback distribution.

The initial conditions are adopted from \citet{Oser2010} with WMAP3 cosmological parameters \citep[][$h=0.72, \;\Omega_{\mathrm{b}}=0.044, \; \Omega_{\mathrm{dm}}=0.26, \;\Omega_{\Lambda}=0.74, \; \sigma_8=0.77 $, and $\mathrm{n_s}=0.95$]{2007ApJS..170..377S}. In this paper we use the fiducial resolution runs of \citet{Choi2017} with the baryonic mass resolution of $m_{*,gas}=5.8 \times 10^{6} \Msun$, and the dark matter particle resolution of $m_{\mathrm{dm}} = 3.4 \times 10^{7} \Msun$. The comoving gravitational softening lengths are $\epsilon_{\mathrm{gas,star}} = 0.556 \rm \, kpc $ for the gas and star particles and $\epsilon_{\mathrm{halo}} = 1.236 \rm \, kpc$ for the dark matter. 

\subsubsection{Star formation and chemical enrichment model}
In the simulation, star formation rate is calculated as $d \rho_{\ast} /dt = \eta \rho_{\rm gas} /t_{\rm dyn}$, where $\rho_{\ast}$, $\rho_{\rm gas}$ and $t_{\rm dyn}$ are the stellar and gas densities, and local dynamical time for gas particle respectively. The star formation efficiency $\eta$ is set to $0.025$. Star particles are stochastically spawned when the gas density exceeds a density threshold, $n_{\rm th} \equiv n_0 \left( T_{\rm gas}/ T_0 \right)^3 \left( M_0 / M_{\rm gas}\right)^2 $, where the critical threshold density and temperature are $n_0 = 2.0 \thinspace \rm cm^{-3}$ and $T_0 = 12000$~K and $M_0$ is the gas particle mass in fiducial resolution. This requires that the gas density should be higher than the value for the Jeans gravitational instability of a mass $M_{\rm gas}$ at temperature  $T_{\rm gas}$. 
Following \cite{2013MNRAS.434.3142A}, chemical enrichment is allowed via three channels, winds driven by Type~I Supernovae (SNe), Type~II SNe and asymptotic giant branch (AGB) stars. The chemical yields are respectively adopted from \citet{1999ApJS..125..439I,1995ApJS..101..181W,2010MNRAS.403.1413K} for Type I, Type II SNe and AGB stars. The masses in 11 different species, H, He, C, N, O, Ne, Mg, Si, S, Ca and Fe, are explicitly traced for star and gas particles. Then, the net cooling rate of gas particles is calculated based on the individual element abundances, along with their temperature and density. The cooling rate is adopted from \citet{2009MNRAS.393...99W} for optically thin gas in ionization equilibrium and a redshift dependent UV/X-ray and cosmic microwave background is adopted from \citet{2001cghr.confE..64H}. Finally, metal enriched gas particles are allowed to mix their metals with neighboring gas via turbulent diffusion of gas-phase metals as in \cite{2013MNRAS.434.3142A}.

\subsubsection{Stellar feedback model}\label{sec:snmodel}
The stellar feedback model is adopted from \citet{2017ApJ...836..204N}. The stellar feedback processes included in the simulations are 1) winds from young massive stars, 2) UV heating within Str$\rm \ddot{o}$mgren spheres of young stars, 3) three-phase Supernova remnant input from both type I and type II SN feedback, and 4) outflow and metals from dying low-mass AGB stars. 

First, the young star particles disperse their wind momentum to nearby gas particles with the same amount of ejected mass and momentum as those of type II SN explosions evenly spread in time before the actual SN event. In addition, their ionizing radiation elevates the temperature of neighboring gas to $T=10^4$~K within a Str\"{o}mgren radius \citep{1939ApJ....89..526S}.

In the ‘snowplow’ SN feedback model \citep{2017ApJ...836..204N}, it is assumed that a single SN event ejects mass in an outflow distributing energy and momentum to the surrounding gas. The fiducial wind velocity is adopted as $v_{\rm out,SN}=4,500 \kms$. 

A special feature of this snowplow SN feedback model is that it assumes that each gas particle is affected by one of the three phases of the SN remnant successive phase, depending on its position from the star particle exploding into the SN. That is, the gas particle receives SN energy in the form of one of the following three phases: (i)~momentum-conserving free expansion phase, (ii)~energy-conserving Sedov-Taylor phase where SN energy is transferred with 70\% as thermal and 30\% as kinetic, and finally (iii)~the snowplow phase where radiative cooling becomes dominant. This SN feedback model launches standard Sedov-Taylor blast-waves carrying energy as 30\% kinetic and 70\% thermal, and both types of energy gradually dissipate with distance from the SN in pressure-driven snowplow phase of SN remnants. 

Finally, old stellar particles still distribute energy, momentum, and metals via AGB winds. The outflowing wind velocity of AGB stars is set to be $v_{\rm out, AGB}=10 \kms$, following the typical observed outflowing velocities of AGB winds \citep[e.g.][]{1992A&amp;AS...93..121N}.

\subsubsection{Black hole formation, growth and feedback model}\label{sec:bhmodel}

In the simulations, new collisionless black hole particles with $1.39 \times 10^5  \Msun$ are seeded at the center of new emerging dark matter halos with mass above $1.39 \times 10^{11} \Msun$. These black holes grow via two channels: mergers with other black holes and gas accretion. Two black holes are allowed to merge when they are close enough within their local SPH smoothing lengths, and their relative velocities are less than the local sound speed. The gas accretion onto a black hole is calculated using Bondi-Hoyle-Lyttleton parameterization \citep{1939PCPS...34..405H,1944MNRAS.104..273B,1952MNRAS.112..195B}. In addition, the soft Bondi criterion is adopted following \cite{Choi2012a}, to prevent the unphysical infall of gas from outside of the Bondi radius. The accretion rate onto the black holes is not artificially capped at the Eddington rate in the simulation. Instead, it includes the Eddington force model that pushes electrons in gas particles outward from the black holes. Therefore super-Eddington gas accretion occasionally happens in the simulation but the corresponding Eddington force, as well as feedback effects listed below naturally reduce the inflow.

The simulation includes mechanical AGN feedback via broad absorption line winds and radiative X-ray feedback via heating and associated radiation pressure.

In the mechanical AGN feedback model, AGN-driven winds are launched in the vicinity of the black hole with a constant velocity $v_{\rm outf,AGN} =10,000 \kms$, and the wind mass is determined by a mass inflow rate and also by feedback efficiency parameter as $\dot{M}_{\rm acc} = \dot{M}_{\rm inf} \frac{1}{1+\psi}$ where the feedback efficiency parameter $\epsilon_{w}$ is set to 0.005 \citep{Choi2017}. The total energy flux carried by the wind is $ \dot{E}_w \equiv \epsilon_w \dot{M}_{\rm acc} c^2=0.5 \dot{M}_{\rm outf} v_{\rm outf,AGN}^2$. The selected feedback efficiency $\epsilon_{w}$ and the AGN wind velocity $v_{\rm outf,AGN}$ correspond to the wind model that 90~\% of the inflowing mass is expelled as winds ($\dot{M}_{\rm outf} = 9 \dot{M}_{\rm acc}$) carrying momentum flux of $\dot{p} = 30 L_{\rm BH} / c$. The wind is directed to be parallel or anti-parallel to the angular momentum vector of each gas particle accreted by the black hole.

In the radiative X-ray feedback model, the heating and associated radiation pressure effect from moderately hard X-ray radiation ($\sim 50$~keV) from the accreting black hole are included. Using the Compton and photoionization heating prescription for the AGN optically-thin spectrum from \citet{2004MNRAS.347..144S,2005MNRAS.358..168S}, the heating rate is calculated for all gas particles, each affected by all accreting BHs within the simulation. In addition, X-ray radiation pressure directed away from the black hole is also included \citep[see also][]{2011ApJ...738...16H}. 

\subsection{Strong vs. weak stellar feedback}

Supernova explosions eject gas and dust, transferring both thermal and kinetic energy to the surrounding gas. This energy disrupts the interstellar medium, impeding cooling processes and, as a result, suppressing star formation. Supernova feedback has become a crucial element in simulations, as it helps resolve discrepancies between simulated and observed galaxy luminosities \citep[see e.g.][for reviews]{2015ARA&A..53...51S,2017ARA&A..55...59N}. The effects of supernova feedback are more significant in smaller, fainter galaxies than in larger, brighter ones, making satellite galaxies a better subject for studying its impact. In this study, we compare two sets of simulations with different supernova feedback strengths to analyze their influence on the physical properties of satellite galaxies.

The first model is {\bf the fiducial model} that incorporates the three-phase SN winds as described in \autoref{sec:snmodel}, using an outflow velocity of $v_{\rm out,SN}=4,500 \kms$, following the framework established by \citet{2017ApJ...836..204N}. This represents a mechanical supernova (SN) feedback model.  

The second model is {\bf the weak SN feedback model}, identical to the fiducial model except for its reduced outflow velocity of $v_{\rm out,SN}=10 \kms$. Since directly comparing mechanical and thermal SN feedback models under {\it equivalent} conditions is challenging, we adopt the approach of lowering the outflow velocity in the mechanical SN feedback model, rather than using a completely distinct thermal feedback model. This adjustment effectively diminishes the impact of wind emission while retaining the ability of SN ejecta to enrich the surrounding gas with metal-rich material. Consequently, this model behaves as a pseudo-thermal feedback model, where the majority of the energy is released as heat.

\subsection{Halo finders \& Identifying satellites}
Among 30 zoom-in simulated galaxies presented in \citet{Choi2017}, we select 11 galaxies with stellar mass cut of $10^{11} < \Mstel/\Msun < 2 \times 10^{11}$ for central galaxies, following \citet{2015MNRAS.454.1605R}. These galaxies have identification number m0209, m0329, m0380, m0408, m0501, m0549, m0664, m0721, m0763, m0858, and m0908, with virial masses of $1.4 \times 10^{12} \le M_{\mathrm{vir}}/\Msun \le 6.6 \times 10^{12}$.

We identify central host galaxies and their satellite galaxies in all 22 zoom-in simulations, comprising 11 simulations for each of the two different stellar feedback models. We first utilize publicly available software, \texttt{ROCKSTAR} \citep{behroozi2013}, a phase-space temporal halo finder, to obtain the halo catalog. The virial radius and masses of halos are determined using the spherical over-density threshold, $\rho_{200}$, i.e., $M_{\rm vir} \equiv M_{200}$ and $r_{\rm vir}\equiv r_{200}$. Subsequently, we validate galaxy locations with \texttt{pygad} \citep{Rottgers2020a} and calculate the galaxy stellar masses. The stellar mass of simulated galaxies is defined as the total mass of star particles within 10 \% of the virial radius, $r_{10}$.

For each zoom-in simulation, we set the most massive galaxy as the central host galaxy. We find satellite galaxies using the same criteria as \citet{2015MNRAS.454.1605R}: galaxies with a projected distance of less than 300 kpc from the host galaxy and a mass of $10^9 \Msun$ or more are selected as satellite galaxies.

We impose a limit of 64 dark matter particles on the halo catalog data since below 64 particles the mass function fails to converge \citep{2007ApJ...655L..17B}. This criterion gives us confidence on the physical properties of halos and subhalos down to a virial mass of $2.18 \times 10^9$~$\Msun$.

\begin{figure*}
\includegraphics[width=\linewidth]{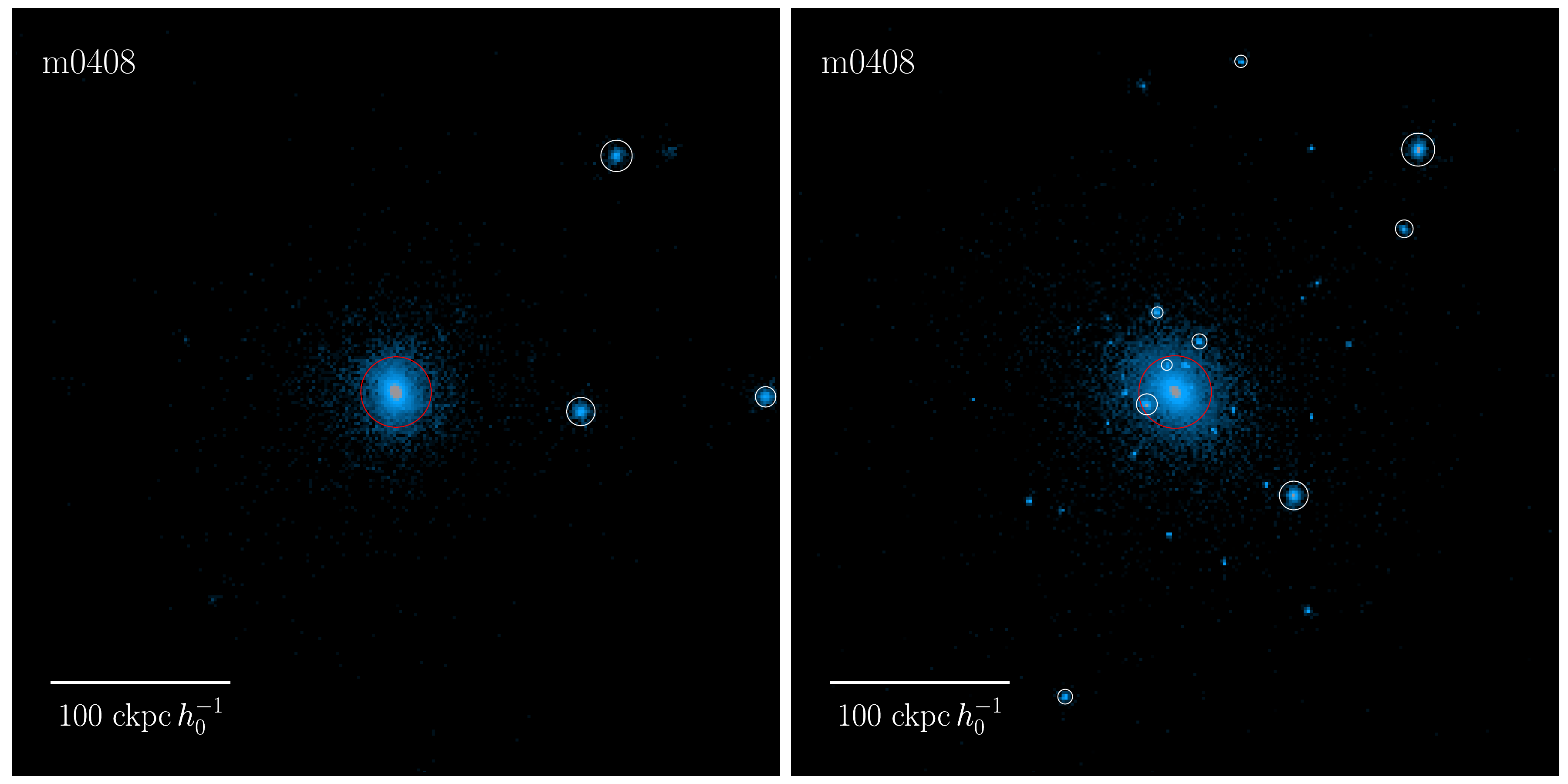}{}
\caption{Galaxy example images of halo m0408 at $z=0$, simulated with fiducial supernova (SN) feedback (left) and weak SN feedback (right). The images were generated using \texttt{pygad}, projected over a cube of 600 kpc on each side. The locations of galaxies identified by ROCKSTAR are marked with circles: central galaxies are highlighted in red, while satellite galaxies with a mass exceeding $10^9 \Msun$ are marked in white. The size of the circles corresponds to the 10 \% of virial radius of the galaxy. }
{\label{stellardensity_408}}
\end{figure*}

\begin{figure*}
\includegraphics[width=\linewidth]{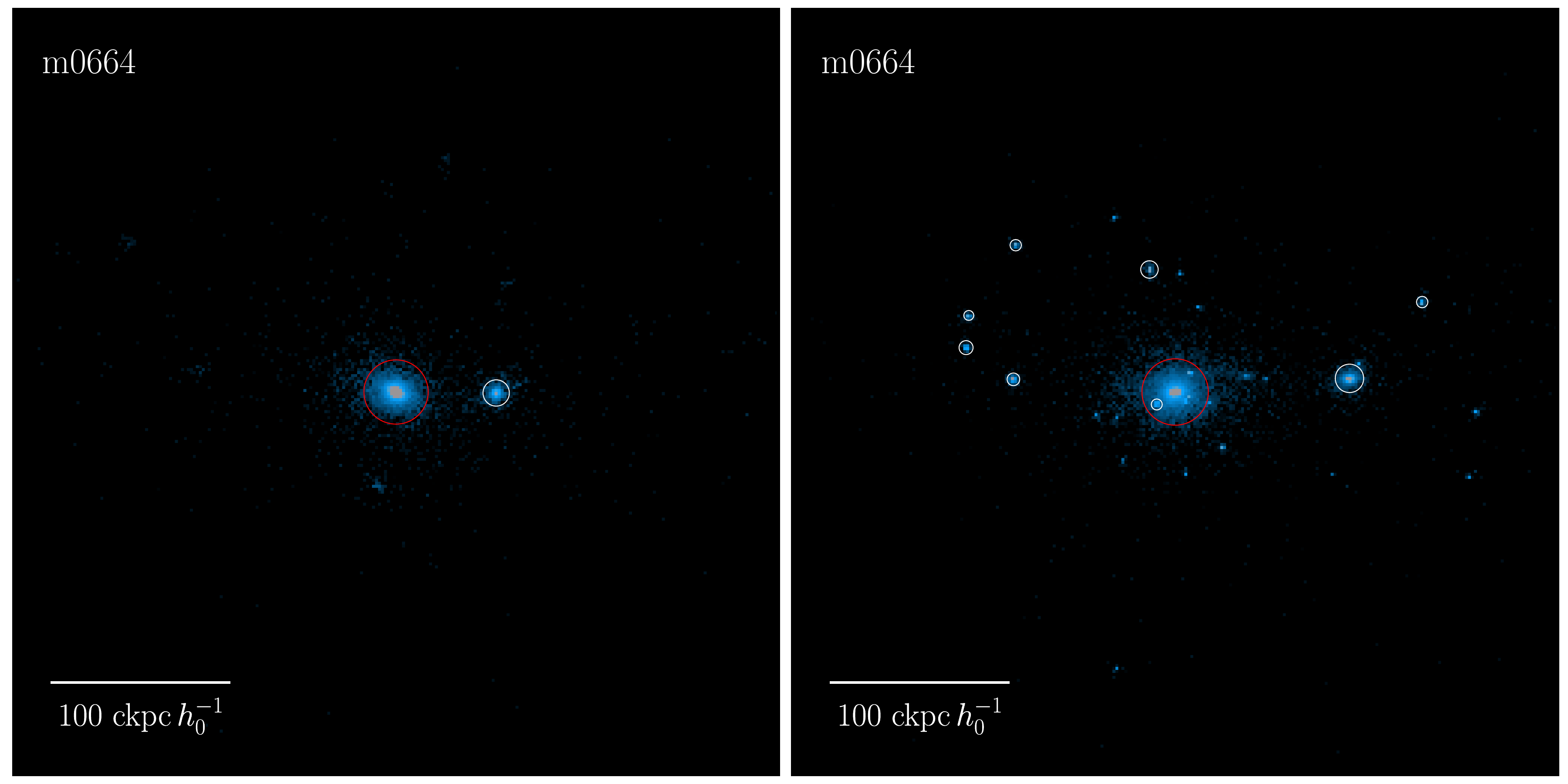}{}
\caption{Same as in \autoref{stellardensity_408}, but for halo m0664. More satellite galaxies are visible in weak SN feedback case shown in right panel.}
{\label{stellardensity_664}}
\end{figure*}

\begin{figure}
\includegraphics[width=\linewidth]{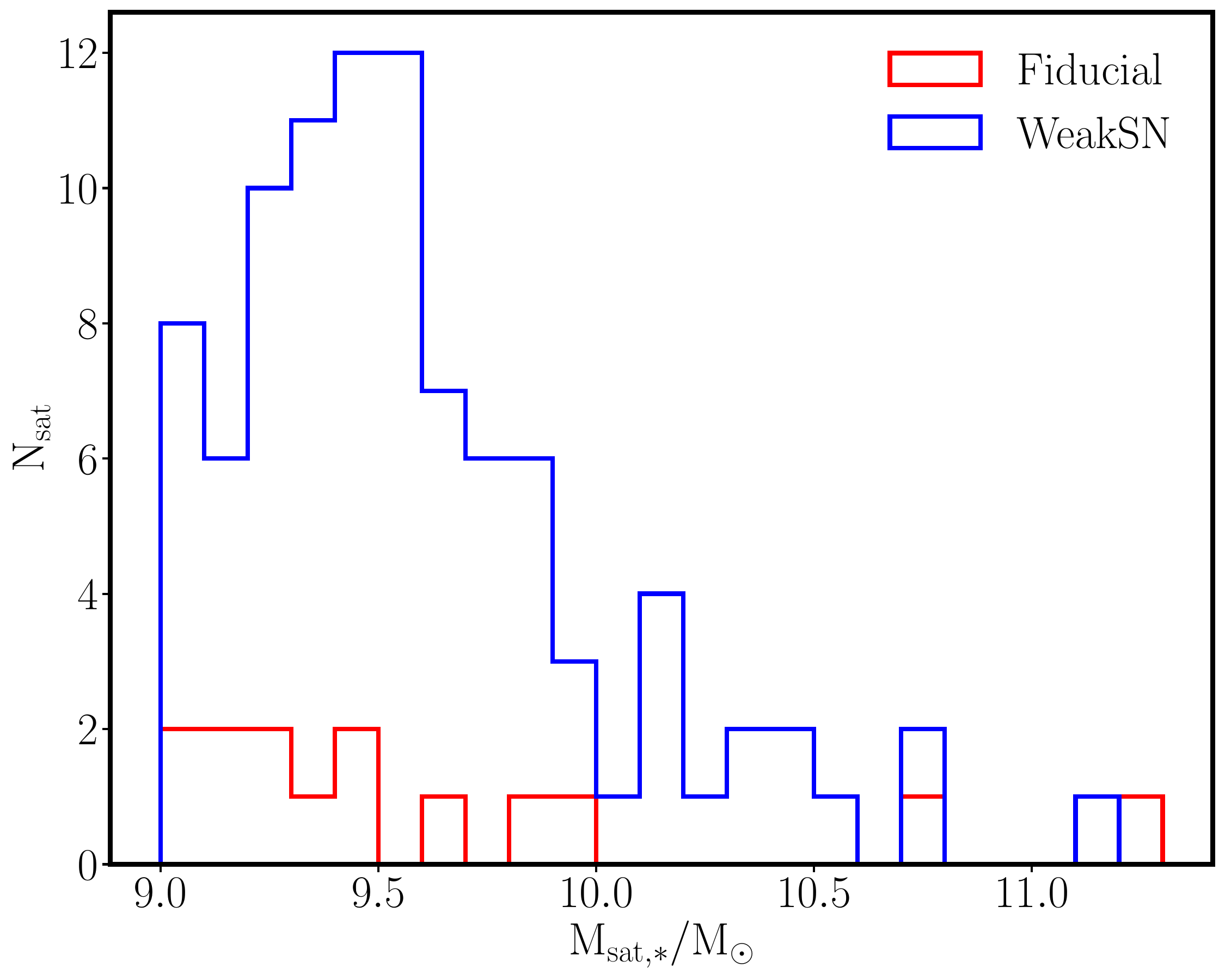}{}
  \caption{Histogram showing the stellar mass distribution of satellite galaxies in fiducial simulations (shown in red) and weak supernova feedback simulations (shown in blue) for 11 zoom-in simulations. Satellite galaxies are selected based on a mass cut of $\Mstel > 10^9 \Msun$ and a projected distance cut of $R_{\rm proj} < 300$ kpc from the central host galaxy.}
  {\label{S_hist}}
\end{figure}

\begin{figure}
    \centering
    \includegraphics[width=\linewidth]{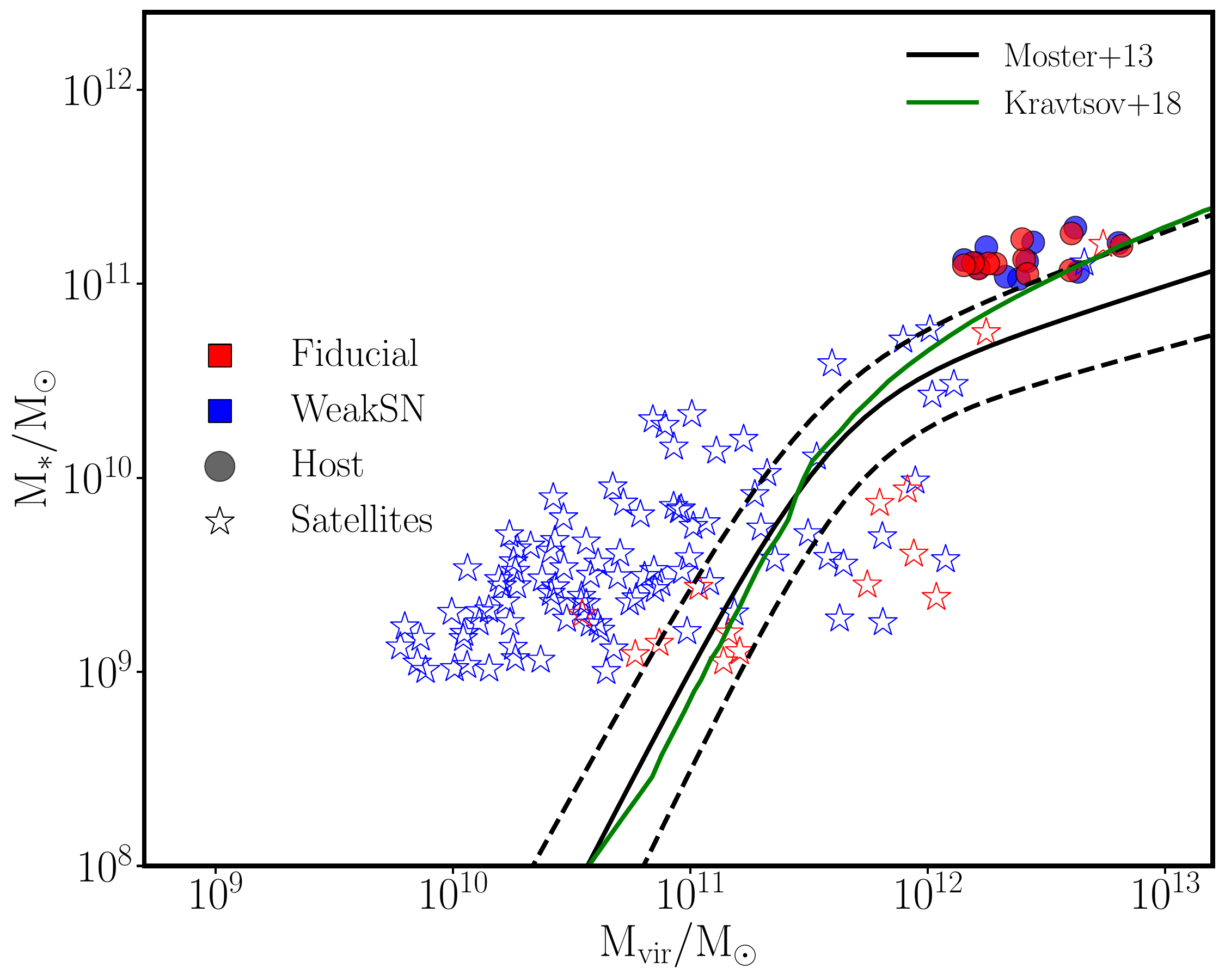}{}
    \caption{Stellar-to-halo mass relation of satellite galaxies and their elliptical host galaxies. The black and green lines represent abundance matching relations from \citet{2013MNRAS.428.3121M} and \citet{Kravtsov2018} (for BCGs), respectively. Dotted lines indicate the corresponding 1$\sigma$ uncertainty ranges. Satellite galaxies used in this analysis are shown as star-shaped markers, applying a stellar mass cut of $\Mstel > 10^9~\Msun$. Host elliptical galaxies are marked with solid circles. Red and blue colors correspond to the Fiducial and Weak SN feedback models, respectively.}
    \label{SMH}
\end{figure}

\begin{figure}
\includegraphics[width=\linewidth]{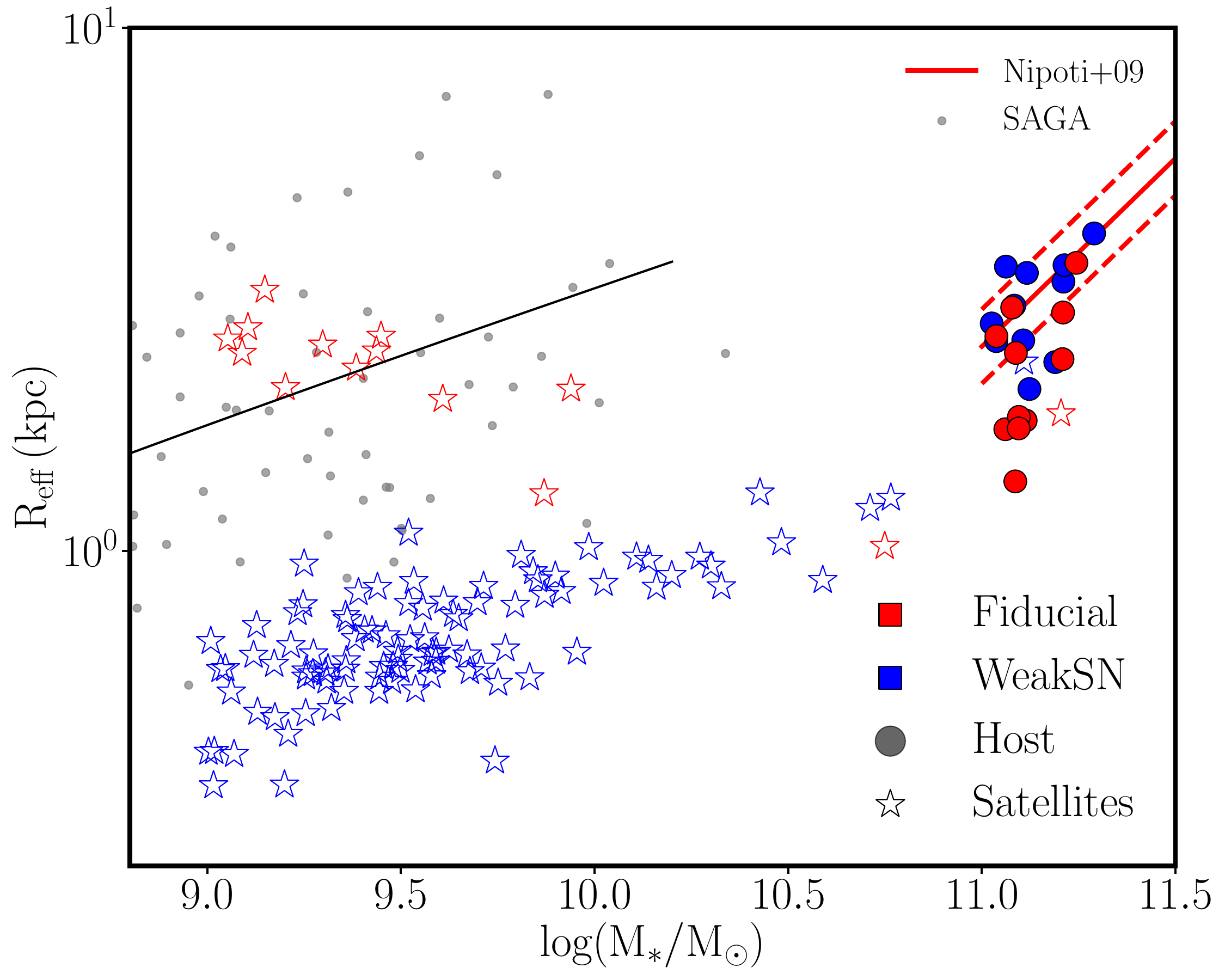}{}
    \caption{Effective radius as a function of stellar mass for central host galaxies (filled circles) and satellite galaxies (open stars) in the fiducial (red) and weak SN feedback (blue) models. The observed size–stellar mass relation for quiescent galaxies from \citet{Nipoti2009} is shown as a solid red line. Observed satellite galaxy sizes from the DR3 SAGA survey \citep{Mao2024} are represented by grey dots, with the black line indicating the fit to the SAGA data. Satellites in the fiducial model have effective radii consistent with the observed SAGA satellite sizes and are approximately 3.5 times larger than those in the weak SN feedback model.}
  {\label{RSM}}
\end{figure}

\section{Result}\label{sec:result}

By examining the differences in the physical properties of satellite galaxies between the fiducial and weak SN feedback models, we investigate the impact of SN feedback on satellite galaxies.

First, we analyze the differences in stellar distributions between the two distinct SN feedback models. \autoref{stellardensity_408} and \autoref{stellardensity_664} show the distributions of both the central host galaxy and its satellite galaxies, of example halos, m0408 and m0664. The central host galaxy stellar masses of m0408 are $1.22 \times 10^{11} \Msun$ and $1.63 \times 10^{11} \Msun$, for fiducial and weak SN feedback, respectively. The central host galaxy stellar masses of m0664 are $1.15 \times 10^{11} \Msun$ and $1.09 \times 10^{11} \Msun$, for fiducial and weak SN feedback, respectively. Galaxies meeting the selection criteria for satellite galaxies, i.e., a projected distance $R_{\rm proj}<300$~kpc and a stellar mass $\Mstel>10^9 \Msun$, are marked with white circles. It is evident that the weak SN feedback model yields approximately 4-5 times more satellites. Furthermore, numerous small stellar clumps which do not meet the satellite mass threshold are present in the halo simulated with weak SN feedback.

\subsection{Satellite mass distribution}\label{Smass}
\autoref{S_hist} displays the stellar mass distribution of a total of 109 satellite galaxies, comprising 95 satellites from the weak SN feedback and 14 satellites from the fiducial model, which pass our stellar mass cut of $\Mstel > 10^9 \Msun$ and distance criteria of $R_{\rm proj}<300$ kpc from the central galaxy. For both SN models, the number of satellites generally decreases with increasing mass, with most satellites having stellar masses below $10^{10} \Msun$. The number of satellites is consistently higher across the stellar mass range for the weak SN feedback model compared to fiducial model. In particular, in the stellar mass range of $10^{9} < \Mstel/\Msun < 10^{10}$, the number of satellites in the weak SN feedback model is notably higher. This feature of mass distribution of satellite galaxies explain differences of galaxy image examples of two sets shown in \autoref{stellardensity_408} and \autoref{stellardensity_664}.

We examine the normalized radial distribution of satellite galaxies in both simulations to assess their spatial distribution. Adopting the same stellar mass threshold as in \autoref{S_hist}, $\Mstel > 10^9 \Msun$, we find that the radial profiles are largely consistent between the two models. Minor differences at small radii appear to arise from limited number statistics. Taken together with the results in Figure 3, we conclude that although the weak SN feedback model produces a significantly larger number of satellites compared to the fiducial model, their spatial distributions remain broadly similar.

\autoref{SMH} illustrates the stellar-to-halo mass relation (SHMR) of central and satellite galaxies from our simulations, compared with empirical abundance matching relations from \citet{2013MNRAS.428.3121M} and \citet{Kravtsov2018}. These reference curves primarily describe central (host) galaxies, and our simulated central galaxies are broadly consistent with these observationally constrained relations. In contrast, satellite galaxies exhibit a wider scatter and systematically higher stellar-to-halo mass ratios, particularly in the Weak SN feedback model.

While central galaxies show relatively similar distributions across both fiducial and weak SN feedback simulations, satellite galaxies demonstrate a clear model-dependent separation. Specifically, satellites in the weak SN feedback simulations tend to have higher stellar masses for a given virial mass, locating well above the empirical abundance matching relations. Conversely, satellites in the fiducial feedback scenario generally have lower stellar masses relative to their halo masses, typically falling below the reference curves.

This distinct difference arises due to the varying strengths of the SN feedback used in our models. The strong SN-driven winds in the fiducial model effectively suppress star formation, whereas the weaker winds in the weak SN feedback model are insufficient for significant suppression, resulting in more massive satellite galaxies at lower virial masses. This effect is particularly noticeable at the virial masses $\Mvir < 10^{11} \Msun$. However, at higher virial masses ($\Mvir \gtrsim 10^{12}\Msun$), the stellar masses of satellites in both feedback models become comparable due to the increasing dominance of AGN feedback.

In \autoref{RSM}, we present the effective radii of host and satellite galaxies as a function of stellar mass. The effective radius is calculated by averaging the half-mass radius projected along three axes. For comparison with observations, we include the observed size–stellar mass relation for massive quiescent galaxies ($10^{11} \le \Mstel/\Msun \le 10^{12}$) from the SLACS survey \citep{Nipoti2009}, as well as the satellite galaxy sizes from Data Release 3 of the SAGA survey \citep{Mao2024}, which studied satellite galaxies around Local Group–mass hosts. The SAGA survey focuses on satellite galaxies around Milky Way–mass host galaxies at very low redshifts ($z < 0.015$), covering distances of 25 to 40.75 Mpc from the Local Group \citep{Geha2017,Geha2024,2024arXiv240414500W}. The survey identifies 378 satellites across 101 host galaxies, which are less massive than our simulated host galaxies. In the SAGA study, satellite galaxies are classified by their magnitude, distance, and heliocentric velocity relative to the host. The stellar masses of confirmed satellite galaxies in the SAGA dataset range $10^6 < \Mstel/\Msun < 10^{10.6}$, comparable to the stellar mass range of our simulated satellite galaxies.
 
This figure illustrates the impact of mechanical SN feedback on the effective radii of both host and satellite galaxies. While the host galaxies show relatively small differences in effective radius between the two models, the satellite galaxies exhibit distinct distributions in the two simulation sets. As shown in \autoref{S_hist}, the weak SN feedback model leads to a higher number of satellites. These galaxies have relatively small effective radii, around 1 kpc. In contrast, the satellite galaxies in the fiducial set exhibit larger effective radii, exceeding 1 kpc, approximately 3.5 times greater than those in the weak SN feedback model, despite having similar stellar mass ranges. This size distribution in the fiducial model is broadly consistent with the SAGA survey results for satellite galaxies.  

To clarify the apparent differences in the stellar mass–size relation of satellite galaxies shown in \autoref{RSM}, we examine their gas fractions and compared the results with observations. In the Weak SN feedback simulation, satellite galaxies exhibit a positive correlation between stellar mass and effective radius, reflecting their predominantly quenched nature and low gas content. This trend is qualitatively consistent with SDSS–MaNGA observations of quenched satellites \citep{Biswas2024}. In contrast, the Fiducial simulation appears to show a weak negative trend, driven by a subset of low-mass satellites with relatively large sizes. However, the overall distribution is in good agreement with the SAGA survey data, and both star-forming and quenched satellites in the Fiducial run fall within the observed range of size–mass relations. We note that the limited number of satellites in the Fiducial model (14 in total) may contribute to the perceived trend, and both simulations remain broadly consistent with available observations when accounting for sample size and feedback model differences. We also note that the target galaxies of the SAGA survey are Milky Way analogs, whereas the host galaxies in our study reside in a higher mass range.

Additionally, in line with our adopted black hole seeding model, dark matter halos with $M_{\mathrm{vir}} > 1.39 \times 10^{11} \Msun$ are seeded with black holes. Galaxies within halos above this mass threshold are significantly influenced by AGN feedback. Therefore, the most massive satellites in both the weak SN feedback and fiducial models have effective radii similar to those of the central host galaxies.

The central host galaxies in the weak SN feedback model also exhibit large effective radii, comparable to the distribution of satellite galaxies in the fiducial model. This similarity is likely driven by the strong influence of AGN feedback, with effective radii values aligning closely with those of the central host galaxies in the fiducial model. We explore this further in the \autoref{sec:discussion}.

\begin{figure}
\includegraphics[width=\linewidth]{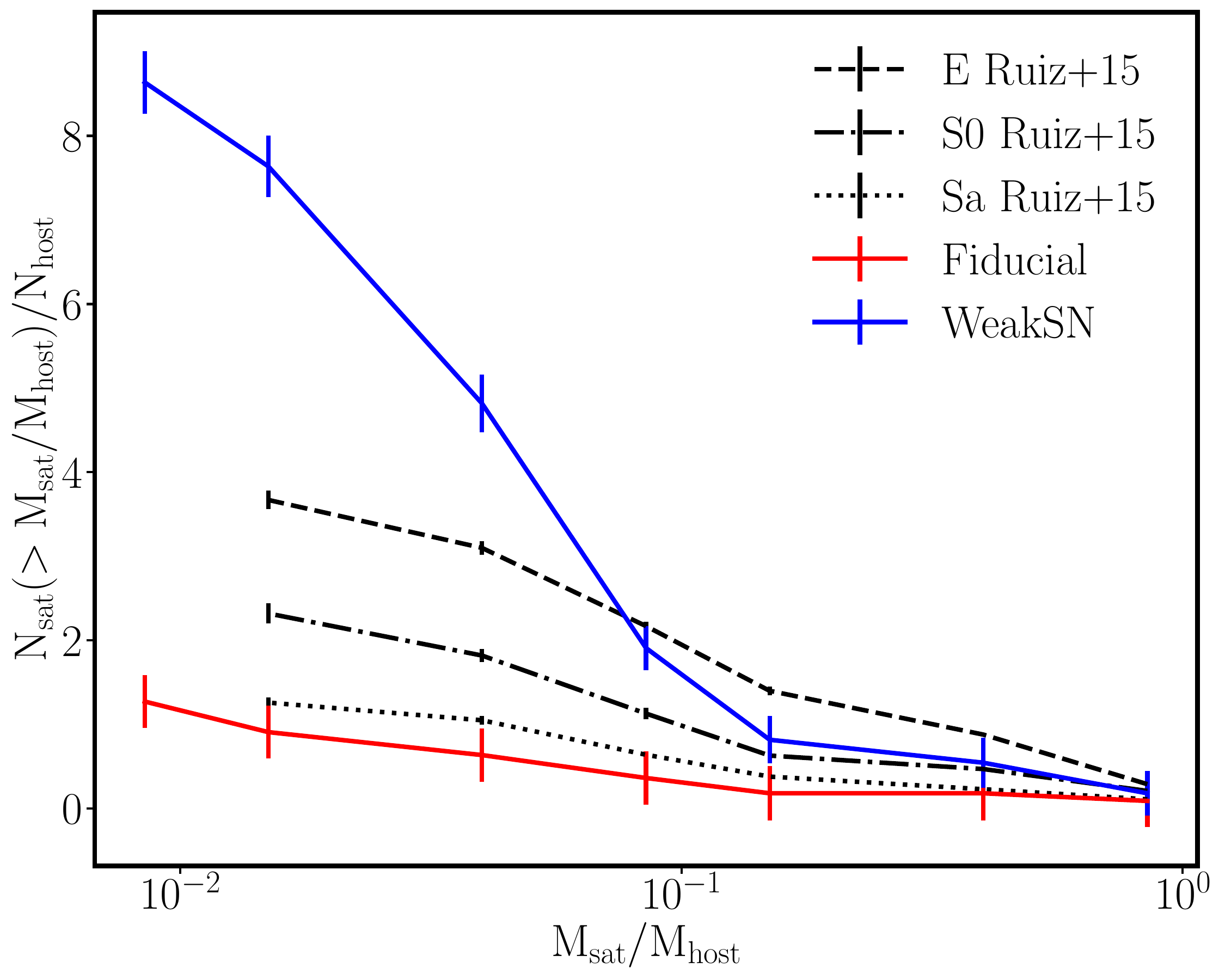}{}
  \caption{Cumulative number of satellites per galaxy host as a function of the stellar mass ratio between satellite and host galaxy, down to 1:100, for fiducial simulations (red) and weak SN feedback simulations (blue). Observed number distributions are obtained from \citet{2015MNRAS.454.1605R}, categorized by the morphological type of host galaxies: elliptical, S0, and Sa. The satellites in the simulation are chosen using the same distance and mass criteria as in \citet{2015MNRAS.454.1605R}. For the weak SN feedback model, a significant number of small-mass satellites are formed.}
  {\label{NM}}
\end{figure}

\subsection{Number of satellites per galaxy host}
Turning our focus, we delve into the abundance of satellite galaxies around elliptical galaxies and compare them to observations. Our investigation involves comparing the abundance of satellite galaxies around elliptical galaxies simulated with two distinct stellar feedback models, followed by a comparison of these findings with observations. 

\autoref{NM} depicts the cumulative number of satellites per galaxy host as a function of the stellar mass ratio between satellite and host galaxy. For comparison, we show observed number distributions from \citet{2015MNRAS.454.1605R}, who investigated the distribution of satellite galaxies around host galaxies of various morphological types in the nearby universe ($z \le 0.025$) using Sloan Digital Sky Survey (SDSS) Data Release 10 (DR10). We compute the stellar mass ratio of $M_{\text{sat},\ast}$ to $M_{\text{host},\ast}$ for all simulated satellites above the mass cut and tabulate the counts within each bin. Considering the stellar mass criteria in our study, the minimum value of $M_{\text{sat}} / M_{\text{host}}$ is 0.005, which is smaller than the lower limit of \citet{2015MNRAS.454.1605R}. Hence, we align our lowest bin value with that of \citet{2015MNRAS.454.1605R}.

Given the prevalence of satellites with low mass ratios in both models, the cumulative number of satellites exhibits a sharp increase towards lower mass ratios. On average, the number of satellites with a mass ratio of 0.1 or greater per host is approximately $\sim1$ for the fiducial model and $\sim1.9$ for the weak SN feedback model. Similarly, the average number of satellites with a mass ratio of 0.005 or larger is approximately $\sim1.8$ for the fiducial model and $\sim8.5$ for the weak SN feedback model. This indicates that the number of satellites in the weak SN feedback model is significantly higher, particularly when considering smaller mass satellites. Overall, the abundance of satellites in the weak supernova feedback model surpasses that of the fiducial set, consistent with the findings in \autoref{stellardensity_408} and \autoref{stellardensity_664}, which demonstrate that the weak supernova model hosts more satellite galaxies than the fiducial model.

Compared to the observations of \citet{2015MNRAS.454.1605R}, the weak SN feedback model exhibits approximately twice as many satellites as the observed number of elliptical-type hosts in the small mass ratio range, representing a significant discrepancy. However, the trend closely aligns with observations when considering only satellites with a mass ratio of $M_{\text{sat}}/M_{\text{host}}>0.1$. 

In contrast, the fiducial model yields a lower number of satellites compared to the observations of elliptical and S0 hosts in the study by \citet{2015MNRAS.454.1605R}, resembling the cumulative number of satellites in Sa-type host galaxies. 

The observations of \citet{2015MNRAS.454.1605R} were based on a selection of 254 host galaxies in a narrow stellar mass range ($ 1.06 \times 10^{11} \le \Mstel/\Msun \le 1.95 \times 10^{11}$) from SDSS DR10. They categorized them into their respective morphological types to determine the abundance of satellite galaxies for each host morphological type. Specifically, they used 83 elliptical galaxies for their analysis. In our simulated data, we have 11 hosts with predominantly elliptical morphology for both the fiducial and weak SN feedback models, indicating that our analysis may be affected by small number statistics.

Moreover, when generating initial conditions for the zoom-in simulation from the large volume dark matter-only simulation, massive host regions with massive satellites were intentionally excluded \citep{Oser2010}. Consequently, our zoom-in simulation represents a sample with a lower number of satellites, particularly massive ones. Considering these factors, the difference between the fiducial model and observations appears reasonable. The impact of this sample selection from the zoom-in region is consistent in the case of weak SN feedback. For weak SN feedback case, while the number of satellites exceeds that of the observations, galaxies that are comparable in mass to the host galaxy or do not exhibit a significant mass difference with it are notably fewer compared to the observations of elliptical hosts.

\subsection{The amount of satellite mass surrounding the galaxy hosts}
\begin{figure}
\includegraphics[width=\linewidth]{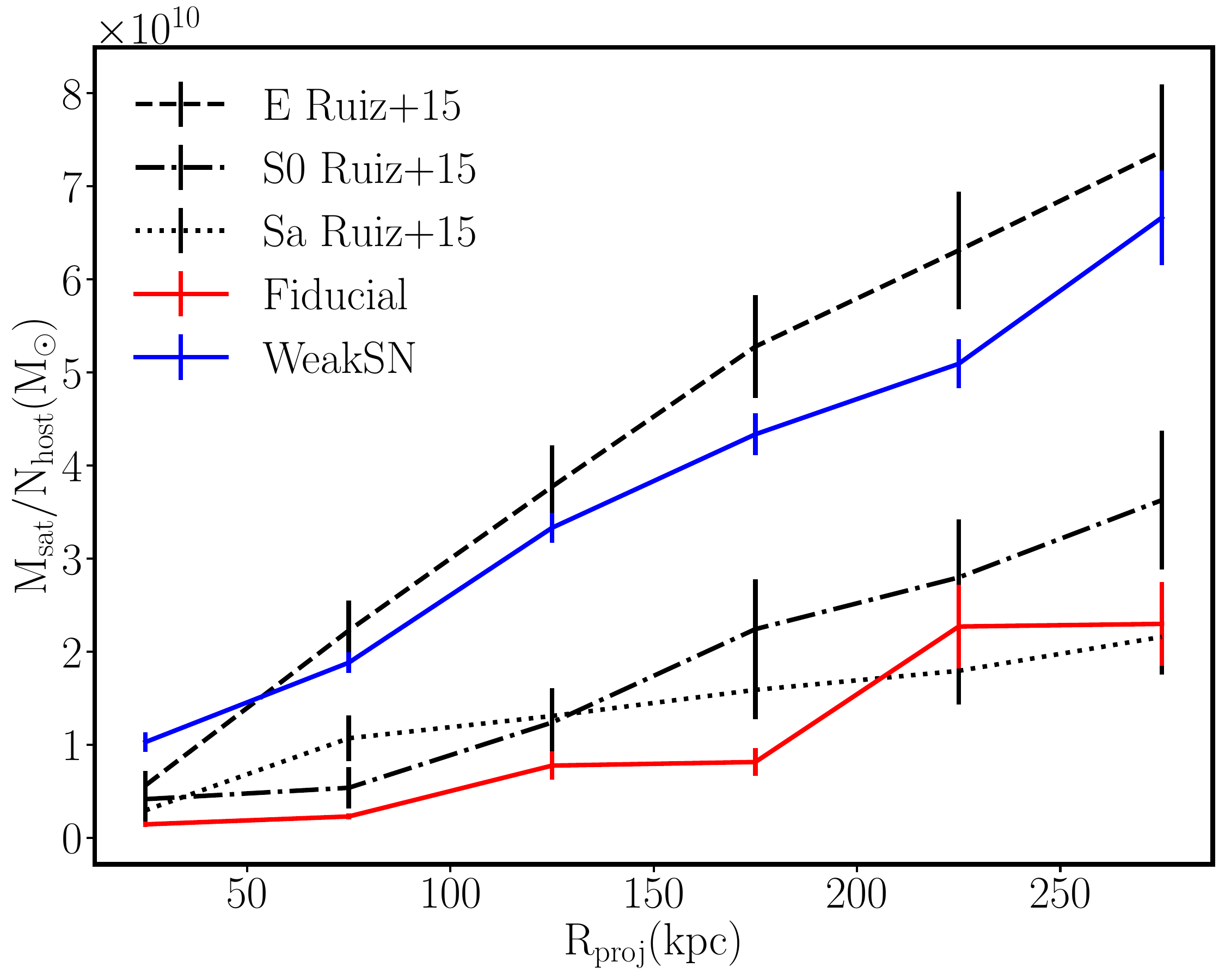}{}
  \caption{
  Cumulative stellar mass enclosed by satellite galaxies per galaxy host down to stellar mass $10^9 \Msun$ as a function of the projected radial distance up to 300 kpc, for fiducial simulations (red) and weak SN feedback simulations (blue). Observed cumulative mass in satellites are from \citet{2015MNRAS.454.1605R}, categorized by the morphological type of host galaxies: elliptical, S0, and Sa. } 
  {\label{MD}}
\end{figure}

In \autoref{MD}, we illustrate the amount of stellar mass accumulated by the satellites around the simulated host galaxies. This calculation extends down to satellites with a stellar mass of $10^9 \Msun$ and is presented as a function of their projected distances from the central host galaxy. Host galaxies simulated with the weak SN feedback are enveloped by approximately three times more stellar mass of satellites compared to fiducial hosts. Notably, in the case of weak SN feedback, the total amount of stellar mass encompassed by the satellites is nearly half the mass of the host galaxy itself (approximately $6 \times 10^{10} \Msun$).

For comparison, we present observational results from \citet{2015MNRAS.454.1605R} for three morphological types of host galaxies (elliptical, S0, and Sa). In the case of weak SN feedback, the cumulative satellite mass tends to closely resemble the observations for elliptical hosts by \citet{2015MNRAS.454.1605R}. As depicted in \autoref{NM}, while the number of satellites in the weak SN feedback model surpasses that of the observations, the masses of the satellites are generally small, resulting in total mass values comparable to the observations. Conversely, in the case of the fiducial model, the cumulative masses closely resemble the observations for Sa hosts or are slightly lower than the observations in the small $R_{\text{proj}}$ range. However, as previously discussed, this discrepancy can be attributed to the exclusion of galaxy hosts with massive satellites during the initial condition generation of the zoom-in simulation. 

\subsection{Satellite number radial profile} 
\begin{figure}
\includegraphics[width=\linewidth]{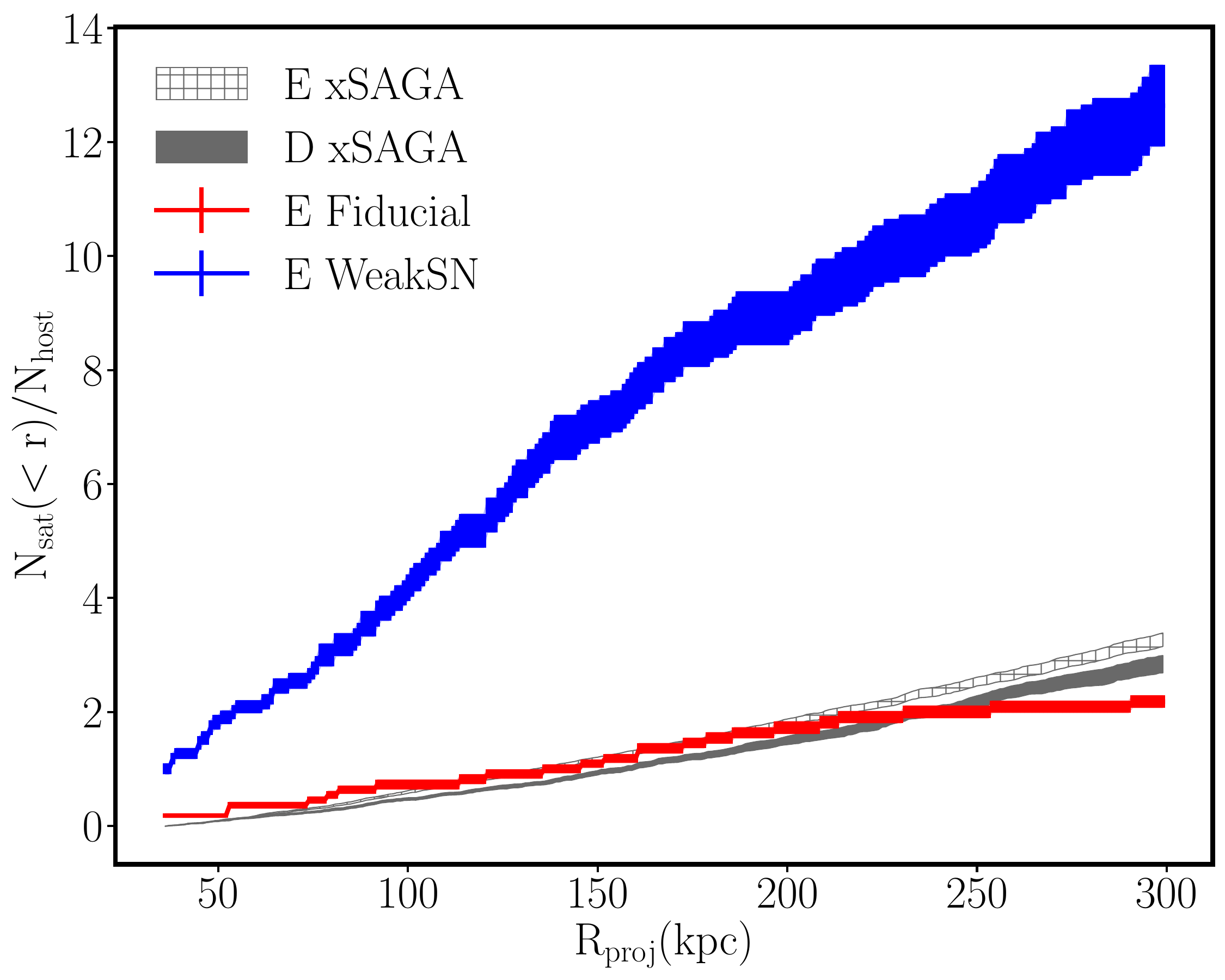}{}
  \caption{
Cumulative number of satellite galaxies as a function of projected radius from the host galaxy, presented for fiducial simulations (red) and weak SN feedback simulations (blue). The observed radial profile is based on eSAGA observations for two morphological types of host galaxies: elliptical and disk \citep{Wu2022}. Following \citet{Wu2022}, satellite galaxies in simulations are selected with a stellar mass cut of $\Mstel > 10^{8} \Msun$. In the case of the weak SN feedback model, a significantly larger number of satellites are formed compared to fiducial model and observation.} 
  {\label{ND}}
\end{figure}

Now, we delve into the radial profile of satellites, examining the average number of satellites as a function of projected radial distance from their hosts. In \autoref{ND}, we present the cumulative number of satellite galaxies as a function of projected radial distance from the host galaxy, considering satellite galaxies down to a stellar mass of $10^8 \Msun$. Note that we reduce the mass cut of the satellite galaxy to $10^8 \Msun$ only for this figure, to align with observational result. Given the resolution of our simulations, adopting a lower stellar mass threshold would yield an insufficient number of stellar particles for robust statistical analysis. Upon examining the number counts down to smaller satellites, we find that within 300 kpc, there are approximately 13 satellites for the weak SN feedback model and about 2 for the fiducial feedback model. While the number of satellites remains relatively stable in the fiducial case compared to the $10^9 \Msun$ mass cut case, it increases by almost 1.5 times in the weak SN feedback model. \autoref{stellardensity_408} and \autoref{stellardensity_664} depict numerous small stellar clumps that did not meet the satellite criteria of $\Mstel > 10^9 \Msun$; their inclusion significantly boosts the number of satellites.

For comparison with observations, we assess the satellite radial distribution of disk and elliptical hosts based on the latest results from the xSAGA (Extending the Satellites Around Galactic Analogs Survey) analysis \citep{Wu2022}. The xSAGA analysis leverages spectroscopic redshift catalogs of satellite populations at distances $D=25-40$ Mpc from the SAGA survey \citep{Geha2017} as a training dataset. Then it utilizes a convolutional neural network to investigate the richness of a larger number of galaxy hosts and their satellite galaxies from the DESI Legacy Imaging Surveys with $z<0.03$. In the xSAGA analysis, a satellite galaxy magnitude limit of $M_r < -15.0$ is adopted, which approximately corresponds to a stellar mass of $\Mstel > 10^{7.5}~\Msun$. Among the results of \citet{Wu2022}, we show the highest mass range (${\rm log}\thinspace( \Mstel/\Msun) = 10.5-11$), which is most comparable to the mass range of our simulated hosts. The weak SN feedback model exhibits significantly richer satellite populations across all radial distance ranges compared to the results from xSAGA. As for the fiducial model, while the shape of the radial profile itself displays some differences, particularly at large radial distances, the satellite counts tend to closely resemble the xSAGA results.

In total, xSAGA analyzed 7542 hosts, and the radial profiles presented in \autoref{ND} utilize 1052 host galaxies for ellipticals and 595 host galaxies for disks. Conversely, we compute radial profiles for 11 elliptical hosts each for fiducial and weak SN feedback models. Consequently, the results of the radial profiles of simulated satellites may be influenced by small number statistics and do not exhibit the same level of smoothness as seen in xSAGA.

\begin{figure}
\includegraphics[width=\linewidth]{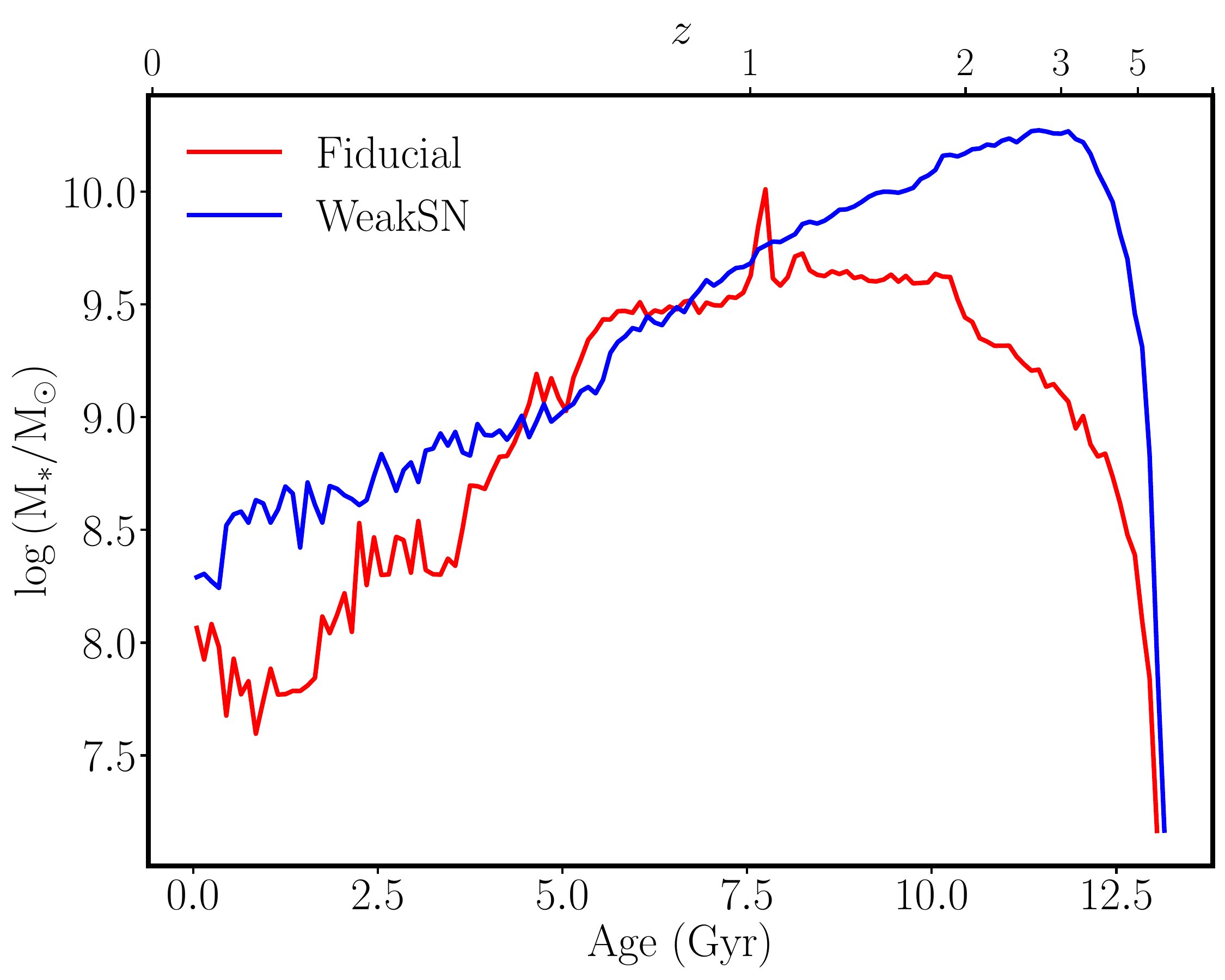}{}
  \caption{Age distribution of all the stars of satellite galaxies with $\Mstel > 10^9 \Msun$, shown for fiducial simulations (red) and weak SN feedback simulations (blue). In the weak SN feedback model, the majority of stars comprising the satellite galaxies formed within the first 2-3 Gyr. Conversely, in the fiducial model, the peak of star formation occurs at a  later time.}
  {\label{AgeM}}
\end{figure}

\subsection{Star formation histories of satellite galaxies} 
We now study the star formation history of the satellites in both simulation sets. In \autoref{AgeM}, we present the age distribution of all the stars constituting the 14 satellites of the fiducial feedback model and the 95 satellites of the weak SN feedback model that meet the stellar mass and projected radial distance criteria. In the case of weak SN feedback, it is evident that the majority of the stellar mass of the satellites is formed before 11 Gyr, corresponding to a bulk production period at $z \sim 3.04$. Conversely, in the fiducial feedback model, star formation in satellite galaxies is considerably more suppressed compared to the weak SN feedback model during the early universe, with most stars forming at a later time. This suggests that the inclusion of mechanical SN feedback in the fiducial model effectively reduces early star formation in small satellites. Note that the small peak observed at $\rm Age=7$ Gyr in the fiducial model corresponds to a star formation burst in the most massive satellite of the fiducial set.

\begin{figure}
\includegraphics[width=\linewidth]{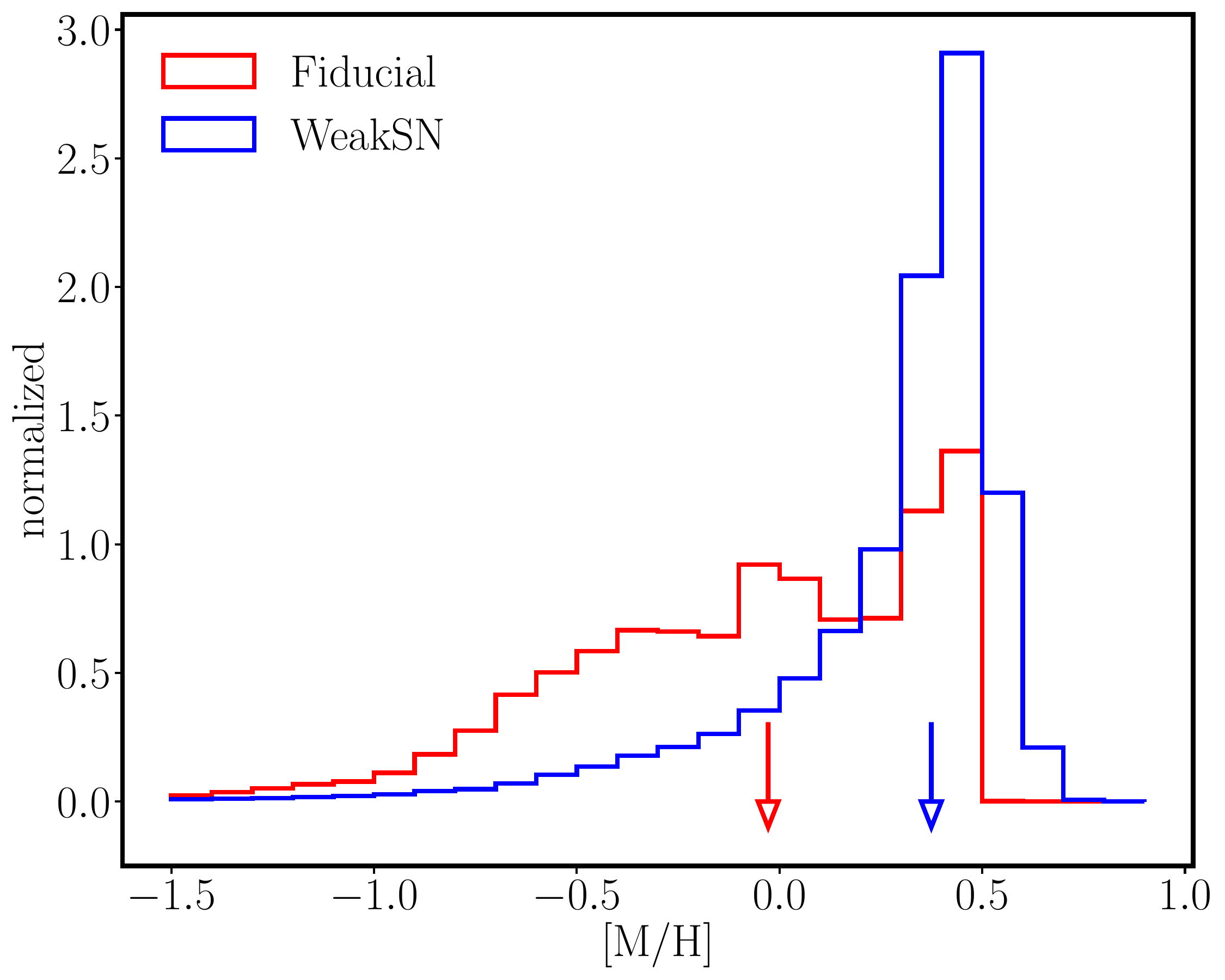}{}
  \caption{
  Metallicity distribution function (MDF) of stars in satellite galaxies with $\Mstel > 10^9 \Msun$, for fiducial simulations (red) and weak SN feedback simulations (blue). Downward arrows indicate the median stellar metallicity values of stars in satellite galaxies in each model. } 
  {\label{MDF}}
\end{figure}

\subsection{Chemical enrichment of satellites}
To compare the chemical enrichment history of satellite galaxies in the two simulation sets, we present the metallicity distribution function (MDF) of all stellar particles comprising satellite galaxies in \autoref{MDF}. Among the 14 satellites in the Fiducial feedback model and the 95 satellites in the weak SN feedback model, we exclude one of the most massive satellites from each model for this analysis as it dominates the MDF distribution. The metallicities of stars in the weak SN feedback model are predominantly super-solar, indicating significant metal enrichment. This phenomenon is attributed to the lack of effective feedback in the Weak SN feedback model, which fails not only to suppress star formation but also to disperse metals away from the vicinity of supernova explosions. As a result, star formation proceeds rapidly in a metal-rich environment, leading to accelerated chemical enrichment and a predominance of metal-rich stellar populations. Conversely, in the fiducial feedback model, stars with super-solar metallicity are also present, but with a broader distribution extending into the metal-poor regime. This distinction is evident when comparing the median metallicities of stars constituting the satellite galaxies in the two models, indicated by the downward arrows. For the fiducial model, the median metallicity is $\rm [M/H]=-0.03$, slightly below solar metallicity, while for the weak feedback model, the median metallicity is $\rm [M/H]=0.37$, reflecting a significant chemical enrichment. 

Since these satellites are analogs of small building blocks that merge into massive elliptical galaxies through minor mergers, the MDF of these satellites has a strong correlation with the MDF of the outer regions of large galaxies. In fact, the metallicity of stars in the outer regions of the central galaxy in the fiducial model is very low, which we have already shown to be consistent with results from resolved stellar photometry observed by the HST \citep{2022ApJ...929..113C}.

\begin{figure}
\includegraphics[width=\linewidth]{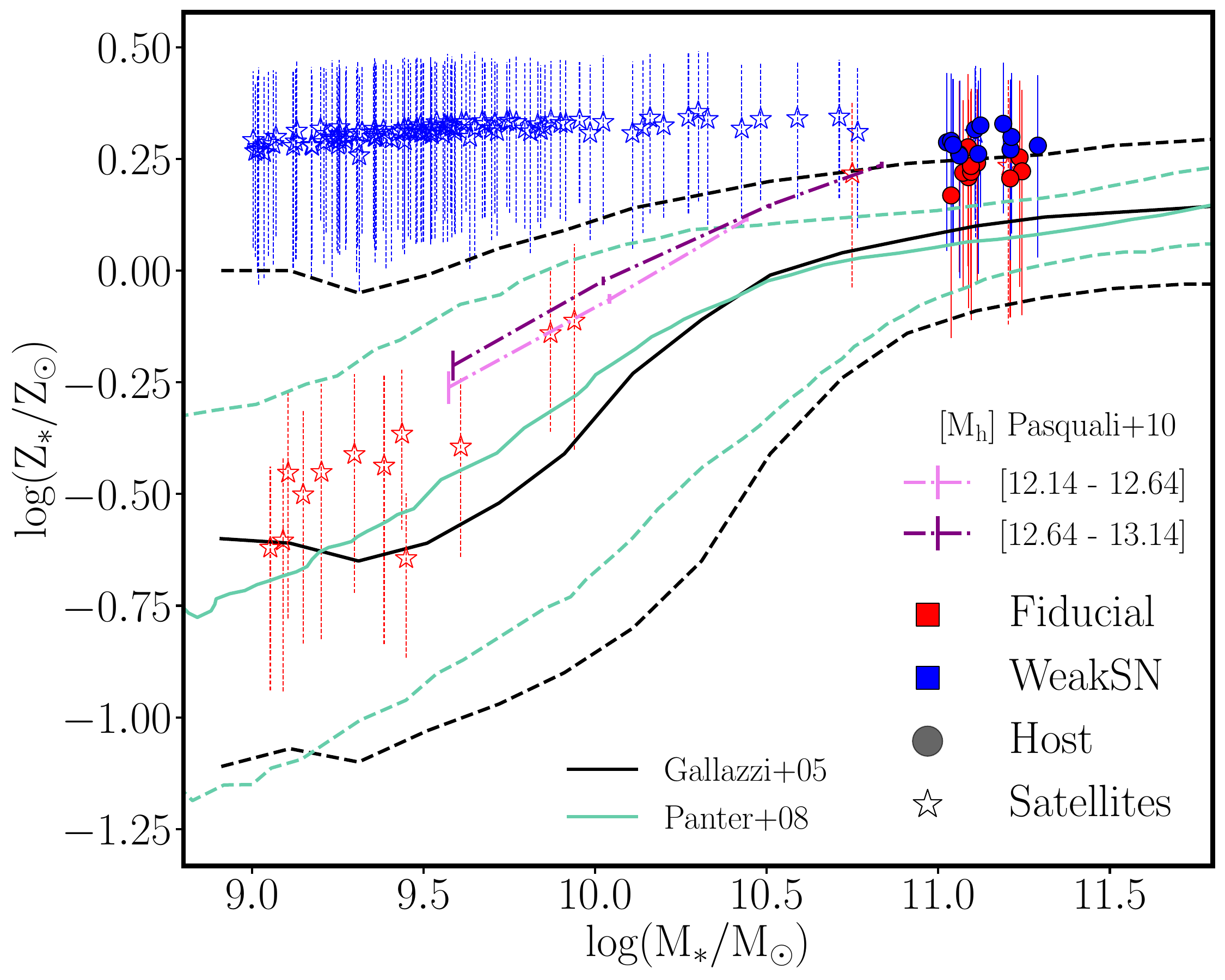}{}
  \caption{
  Mean stellar metallicity as a function of stellar mass (Mass-metallicity relation, MZR) of simulated galaxies in fiducial simulations (red) and weak SN feedback simulations (blue). Host galaxies are shown with filled circles and satellite galaxies with $\Mstel > 10^9 \Msun$ and $R_{\rm proj}<300$ kpc are shown with open stars. The observational MZR from \citet{Gallazzi2005} and \citet{Panter2008} are shown in black and turquoise lines, respectively. Dotted lines indicate the $1 \sigma$ error range for each observational relation. The observed MZR of satellite galaxies is obtained from \citet{Pasquali2010}, represented by pink and purple dot-dashed lines, respectively, for host virial masses ${\rm log } \thinspace (M_{\rm vir}/\Msun)=12.14-12.64$ and $12.64-13.14$, which are close to the mass range of the simulated host galaxies. } 
  {\label{MZR}}
\end{figure}

Now, we study the mean stellar metallicity of each satellite galaxy. In \autoref{MZR}, we show the Mass-Metallicity Relation (MZR) of simulated satellite and host galaxies, i.e., the mean stellar metallicity against their stellar masses. Additionally, we present observed MZRs from \citet{Gallazzi2005} and \citet{Panter2008}, alongside the MZR observations of satellite galaxies from \citet{Pasquali2010} for comparison.

We first examine the metallicities of the host galaxies: those in the weak SN feedback model exhibit slightly higher metallicities compared to the host galaxies of the fiducial model. Nevertheless, the difference is modest, approximately $\sim 0.1$ dex. While the host galaxies of the fiducial model slightly exceed the observed MZRs in terms of metallicities, the difference remains within the observed one sigma range.

However, the difference in the mean stellar metallicity of satellite galaxies between the fiducial and weak SN feedback models is significant. Satellite galaxies in the weak SN feedback model predominantly demonstrate super-solar metallicity values surpassing ${\rm log}\thinspace(Z_{\ast}/Z_{\odot})=0.25$. At $\Mstel/\Msun \sim 10^9$, this marks a substantial deviation of approximately $0.8-1.0$ dex from observations. High mean stellar metallicity of satellite galaxies arises from the rapid and vigorous star formation in the early universe within the small mass halos constituting the satellites in the weak SN feedback model, owing to the absence of effective star formation feedback. Furthermore, in weak SN feedback models with intentionally reduced SN wind velocity, the output from SN explosions remains within the halo potential, chemically enriching the remaining gas quickly. Consequently, stars comprising the satellites become highly metal-rich.

Conversely, in the fiducial model, all satellite galaxies exhibit values consistent with the observed MZR, showing a linear relation between mass and metallicity similar to observations. Notably, satellite galaxies within the stellar mass range studied by \citet{Pasquali2010}, who observed the MZR of satellite galaxies, closely align with the observed MZR line.

\subsection{Gas fraction of satellites}

\begin{figure}
    \centering
    \includegraphics[width=\linewidth]{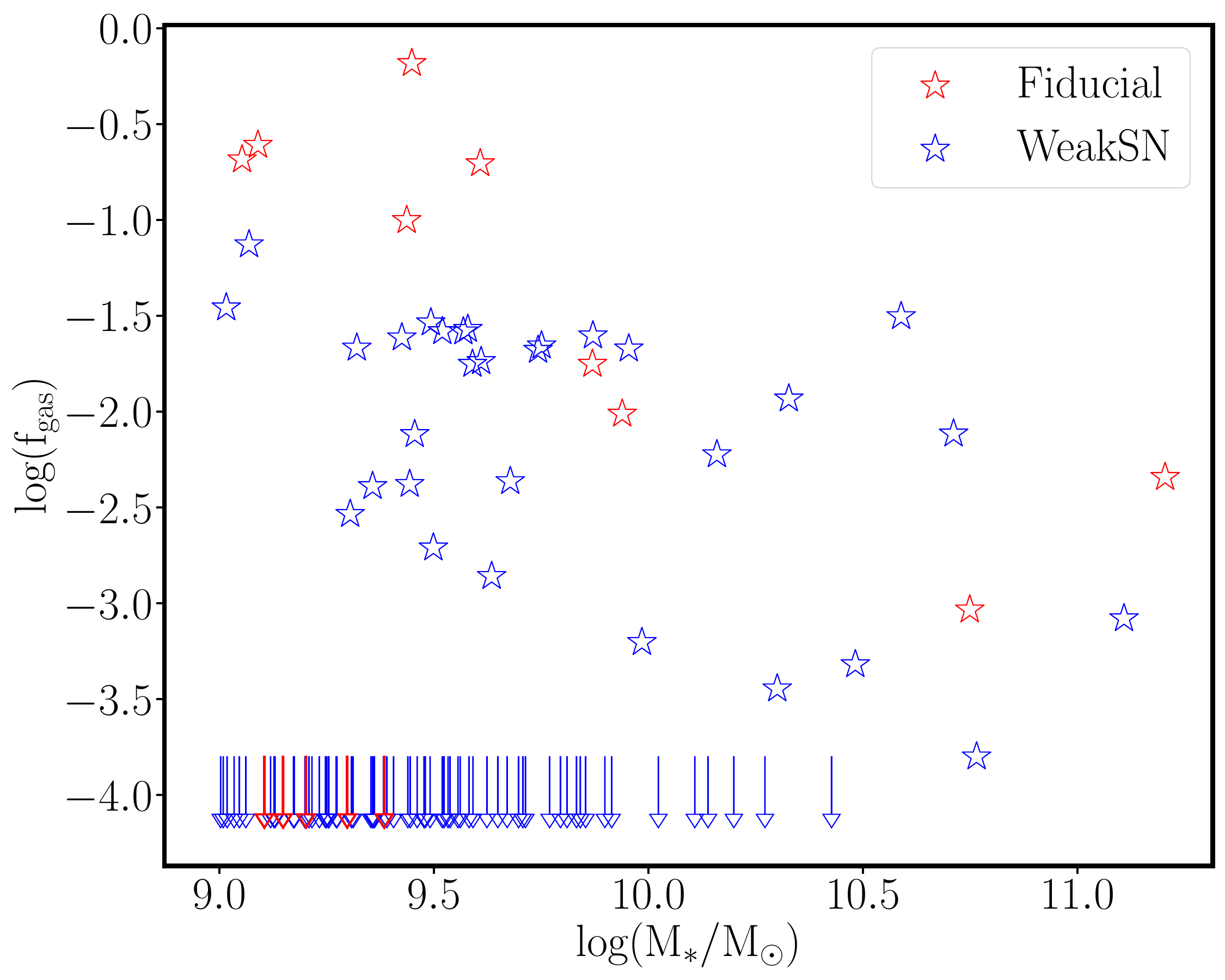}
    \caption{Gas fraction as a function of stellar mass for satellite galaxies in the Fiducial (red) and Weak SN feedback (blue) models. Overall, satellites in the Fiducial simulation exhibit higher gas fractions than those in the Weak SN feedback simulation, particularly at lower stellar masses. Gas-free satellite galaxies are marked with downward arrows in both models.}
    \label{sat_gasf}
\end{figure}

\autoref{sat_gasf} presents the gas fractions of satellite galaxies in both the Fiducial and Weak SN feedback models at $z = 0$. In the Fiducial model, 9 out of 14 satellites retain non-zero gas fractions, whereas in the Weak SN model, only 30 out of 95 satellites contain any gas. This indicates that 35.7\% of satellites in the Fiducial model and 68.4\% in the Weak SN model are gas-free at the present epoch.

Overall, satellite galaxies in the Fiducial model tend to have higher gas fractions than those in the Weak SN model, especially at lower stellar masses. Satellites in the Fiducial run typically show gas fractions in the range of 20–60\%, while most satellites in the Weak SN run remain below 10\%. For instance, the maximum logarithmic gas fraction ${\rm \thinspace log}f_{\rm gas}$ in the Fiducial model is \text{-}0.18, compared to \text{-}1.13 in the Weak SN model.

\section{Discussion and Summary}\label{sec:discussion}

This paper investigates the impact of mechanical supernova (SN) feedback through SN-driven winds on satellite galaxies in massive hosts. Our findings are summarized as follows: 

\begin{enumerate}
    \item {\it Reduction in satellite galaxy number counts}: Mechanical SN feedback significantly reduces the number of satellite galaxies, particularly inhibiting the formation of galaxies with stellar masses smaller than $10^{10} \Msun$ (Figure~\ref{S_hist}, \ref{SMH}, \ref{NM} and \ref{ND}). This effect has already been demonstrated in previous simulations of Milky Way-sized halos \citep[e.g.][]{2010MNRAS.402.1995M}.
    \item {\it Suppression of early star formation}: Mechanical SN feedback effectively suppresses early star formation in satellite galaxies, as reflected in their lower star formation rates (\autoref{AgeM}). Consequently, the total stellar mass of these satellites is also reduced (\autoref{MD}).
    \item{\it Delayed chemical evolution}: In the absence of strong mechanical SN feedback, satellite galaxies undergo rapid chemical enrichment due to active star formation (\autoref{MDF}). Mechanical SN feedback delays this process, resulting in metal-poor satellite galaxies that are in better agreement with observed values (\autoref{MZR}), a phenomenon that has also been discussed in the literature \citep{2007ApJ...655L..17B,2024MNRAS.527.3276I}.
    \item{\it Increased galaxy size}: Simulations with mechanical SN feedback show a notable increase in the effective radius of satellite galaxies (\autoref{RSM}). By suppressing intense star formation in their central regions and redistributing mass through SN winds, these satellite galaxies expand in size, thereby avoiding the overly compact structures previously discussed in the literature \citep[e.g.][]{2008MNRAS.389.1137S,2016ApJ...824...79A}.
\end{enumerate}

The number and mass distribution of satellite galaxies have been extensively studied through both observations and simulations, serving as a critical test for $\Lambda$CDM cosmological models \citep[e.g.][]{2010MNRAS.406..744C,2013ApJ...772..146N}. This study focuses on the influence of mechanical SN feedback via SN-driven winds on satellite galaxies, demonstrating that this feedback mechanism plays a key role in shaping the satellite population. However, many other physical processes also contribute to satellite formation, evolution, and abundance distribution. For instance, \citet{2016ApJ...818..193T,2016ApJ...827L..15T} showed that tidal interactions between the central host and its satellites significantly impact satellite mass loss and redistribution, leading to stripped and deformed satellite structures. Moreover, \citet{Costa2019} revealed that stellar radiation can reduce the number of satellites by as much as 60\%, primarily through the suppression of gas accretion and cooling in lower-mass systems. Additional factors, such as ram pressure stripping \citep{2018MNRAS.478..548S,2024ApJ...960...54Z} and environmental quenching \citep{2015ApJ...808L..27W,2022MNRAS.514.5276S}, further highlight the complexity of the physical processes governing satellite galaxies.

Many of the studies mentioned above have primarily focused on the formation of satellite galaxies within MW–sized halos. This study, however, offers insights by examining satellite formation around group-scale, massive elliptical galaxies and comparing the results to recent observations by \citet{2015MNRAS.454.1605R} and the xSAGA survey \citep{Wu2022}. While the overall mass of the host galaxy could be reproduced similarly to observations due to the strong influence of AGN feedback, even without mechanical SN winds, we find that mechanical SN feedback is essential for accurately reproducing the physical properties of satellite galaxies around the host. This highlights the critical role of SN feedback in shaping satellite galaxy populations in group-scale environments.

The physical properties of massive host galaxies are also influenced by SN feedback, as they are formed through multiple mergers with neighboring satellite galaxies. \citet{Choi2017} demonstrated how the in-situ and ex-situ fractions of stars within the host galaxy vary depending on the strength of SN and AGN feedback models. In the fiducial model with mechanical SN feedback, the average ex-situ star fraction was 60\%, while in the thermal SN feedback model, this fraction increased to 82\% (see Figure 9 in \citet{Choi2017}). This indicates that weaker SN feedback leads to the formation of more massive, and an increased number of satellite galaxies, and more accretion of these satellites into the central galaxy. This effect is indirectly reflected in \autoref{RSM}, where the host galaxies in the weak SN feedback model are larger than those in the fiducial model. The weak SN feedback produces more small satellite galaxies, leading to an increased number of dry minor mergers, as can be seen at \autoref{sat_gasf}, that accrete stars onto the outer regions of the central galaxy. This process expands the stellar envelope and increases the overall size of host galaxy.

\begin{acknowledgments}
We thank the anonymous referee for their constructive and insightful comments, which significantly improved the clarity of this manuscript. We thank Alejandro N\'u\~nez for helpful conversations and suggestions. We also extend our gratitude to John Wu and Yao-Yuan Mao for providing the SAGA survey data.

This work was supported by the National Research Foundation of Korea (NRF) grant funded by the Korea government (MSIT) (No. RS-2025-00515276) for SK and EC. 
TN acknowledges the support of the Deutsche Forschungsgemeinschaft (DFG, German Research Foundation) under Germany’s Excellence Strategy - EXC-2094 - 390783311 of the DFG Cluster of Excellence ``ORIGINS''.
\end{acknowledgments}

\appendix

\section{Test about intermediate outflow velocity}

\begin{figure}
    \centering
    \includegraphics[width=0.7\columnwidth]{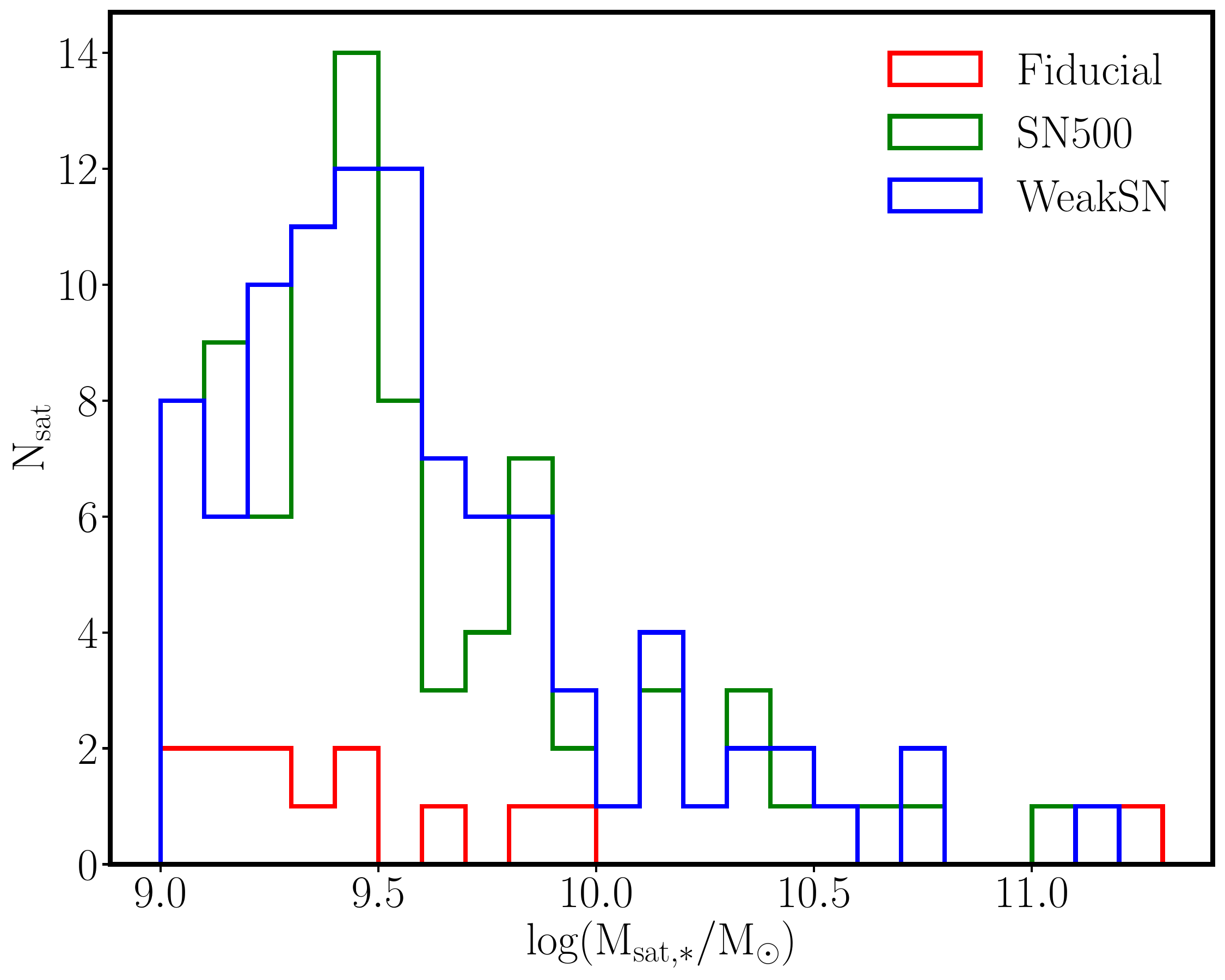}
    \caption{Histogram showing the stellar mass distribution of satellite galaxies from 11 zoom-in simulations. The figure compares results from the fiducial simulation (red), moderate SN feedback simulation with an outflow velocity of 500 $\kms$ (green), and weak SN feedback simulation (blue). The mass and distance selection criteria for satellite galaxies are identical to those described in \autoref{S_hist}.}
    \label{S_hist_wsn500}
\end{figure}

Given that the two SN feedback models in our study represent fiducial (4500 $\kms$) and very weak (10 $\kms$) outflow velocities, we conducted an additional simulation with a moderate outflow velocity of 500 $\kms$ to provide a more systematic test. \autoref{S_hist_wsn500} compares satellite galaxies from the moderate SN feedback model (``SN500'') with those from our two original simulations.

Within the same mass and distance criteria as \autoref{S_hist}, the SN500 model yields 85 satellite galaxies, a number similar to that produced by the weak SN feedback model. Furthermore, the stellar mass distribution of satellites in the moderate feedback simulation closely resembles that of the weak SN feedback case, particularly displaying a significant satellite population in the stellar mass range $10^{9} < \Mstel/\Msun < 10^{10}$.

These findings highlight the critical role of accurately modeling SN feedback in galaxy formation simulations, as even intermediate-strength feedback (500 $\kms$) results in an overabundance of satellite galaxies, similar to the weak SN feedback scenario.

\bibliography{library2}

\begin{thebibliography}{}
\expandafter\ifx\csname natexlab\endcsname\relax\def\natexlab#1{#1}\fi

\bibitem[{{Agertz} \& {Kravtsov}(2016)}]{2016ApJ...824...79A}
{Agertz}, O., \& {Kravtsov}, A.~V. 2016, \apj, 824, 79

\bibitem[{{Agertz} {et~al.}(2013){Agertz}, {Kravtsov}, {Leitner}, \& {Gnedin}}]{2013ApJ...770...25A}
{Agertz}, O., {Kravtsov}, A.~V., {Leitner}, S.~N., \& {Gnedin}, N.~Y. 2013, \apj, 770, 25

\bibitem[{Agertz {et~al.}(2007)Agertz, Moore, Stadel, Potter, Miniati, Read, Mayer, Gawryszczak, Kravtsov, Åke Nordlund, Pearce, Quilis, Rudd, Springel, Stone, Tasker, Teyssier, Wadsley, \& Walder}]{Agertz2007}
Agertz, O., Moore, B., Stadel, J., {et~al.} 2007, Monthly Notices of the Royal Astronomical Society, 380, 963

\bibitem[{Aumer {et~al.}(2013)Aumer, White, Naab, \& Scannapieco}]{2013MNRAS.434.3142A}
Aumer, M., White, S.~D., Naab, T., \& Scannapieco, C. 2013, Monthly Notices of the Royal Astronomical Society, 434, 3142

\bibitem[{{Bechtol} {et~al.}(2015){Bechtol}, {Drlica-Wagner}, {Balbinot}, {Pieres}, {Simon}, {Yanny}, {Santiago}, {Wechsler}, {Frieman}, {Walker}, {Williams}, {Rozo}, {Rykoff}, {Queiroz}, {Luque}, {Benoit-L{\'e}vy}, {Tucker}, {Sevilla}, {Gruendl}, {da Costa}, {Fausti Neto}, {Maia}, {Abbott}, {Allam}, {Armstrong}, {Bauer}, {Bernstein}, {Bernstein}, {Bertin}, {Brooks}, {Buckley-Geer}, {Burke}, {Carnero Rosell}, {Castander}, {Covarrubias}, {D'Andrea}, {DePoy}, {Desai}, {Diehl}, {Eifler}, {Estrada}, {Evrard}, {Fernandez}, {Finley}, {Flaugher}, {Gaztanaga}, {Gerdes}, {Girardi}, {Gladders}, {Gruen}, {Gutierrez}, {Hao}, {Honscheid}, {Jain}, {James}, {Kent}, {Kron}, {Kuehn}, {Kuropatkin}, {Lahav}, {Li}, {Lin}, {Makler}, {March}, {Marshall}, {Martini}, {Merritt}, {Miller}, {Miquel}, {Mohr}, {Neilsen}, {Nichol}, {Nord}, {Ogando}, {Peoples}, {Petravick}, {Plazas}, {Romer}, {Roodman}, {Sako}, {Sanchez}, {Scarpine}, {Schubnell}, {Smith}, {Soares-Santos}, {Sobreira}, {Suchyta}, {Swanson}, {Tarle}, {Thaler}, {Thomas},
  {Wester}, {Zuntz}, \& {DES Collaboration}}]{2015ApJ...807...50B}
{Bechtol}, K., {Drlica-Wagner}, A., {Balbinot}, E., {et~al.} 2015, \apj, 807, 50

\bibitem[{Behroozi {et~al.}(2013)Behroozi, Wechsler, \& Wu}]{behroozi2013}
Behroozi, P.~S., Wechsler, R.~H., \& Wu, H.-Y. 2013, The Astrophysical Journal, 762, 109

\bibitem[{{Belokurov} {et~al.}(2006){Belokurov}, {Zucker}, {Evans}, {Wilkinson}, {Irwin}, {Hodgkin}, {Bramich}, {Irwin}, {Gilmore}, {Willman}, {Vidrih}, {Newberg}, {Wyse}, {Fellhauer}, {Hewett}, {Cole}, {Bell}, {Beers}, {Rockosi}, {Yanny}, {Grebel}, {Schneider}, {Lupton}, {Barentine}, {Brewington}, {Brinkmann}, {Harvanek}, {Kleinman}, {Krzesinski}, {Long}, {Nitta}, {Smith}, \& {Snedden}}]{2006ApJ...647L.111B}
{Belokurov}, V., {Zucker}, D.~B., {Evans}, N.~W., {et~al.} 2006, \apjl, 647, L111

\bibitem[{{Benson} {et~al.}(2003){Benson}, {Bower}, {Frenk}, {Lacey}, {Baugh}, \& {Cole}}]{2003ApJ...599...38B}
{Benson}, A.~J., {Bower}, R.~G., {Frenk}, C.~S., {et~al.} 2003, \apj, 599, 38

\bibitem[{{Biswas} \& {Wadadekar}(2024)}]{Biswas2024}
{Biswas}, P., \& {Wadadekar}, Y. 2024, \apj, 970, 83

\bibitem[{Bondi(1952)}]{1952MNRAS.112..195B}
Bondi, H. 1952, Monthly Notices of the Royal Astronomical Society, 112, 195

\bibitem[{Bondi \& Hoyle(1944)}]{1944MNRAS.104..273B}
Bondi, H., \& Hoyle, F. 1944, Monthly Notices of the Royal Astronomical Society, 104, 273

\bibitem[{{Booth} \& {Schaye}(2009)}]{2009MNRAS.398...53B}
{Booth}, C.~M., \& {Schaye}, J. 2009, \mnras, 398, 53

\bibitem[{{Booth} \& {Schaye}(2013)}]{Booth2013}
---. 2013, Scientific Reports, 1738

\bibitem[{{Brooks} {et~al.}(2007){Brooks}, {Governato}, {Booth}, {Willman}, {Gardner}, {Wadsley}, {Stinson}, \& {Quinn}}]{2007ApJ...655L..17B}
{Brooks}, A.~M., {Governato}, F., {Booth}, C.~M., {et~al.} 2007, \apjl, 655, L17

\bibitem[{{Brooks} {et~al.}(2013){Brooks}, {Kuhlen}, {Zolotov}, \& {Hooper}}]{2013ApJ...765...22B}
{Brooks}, A.~M., {Kuhlen}, M., {Zolotov}, A., \& {Hooper}, D. 2013, \apj, 765, 22

\bibitem[{Brooks \& Zolotov(2014)}]{2014ApJ...786...87B}
Brooks, A.~M., \& Zolotov, A. 2014, The Astrophysical Journal, 786, 87

\bibitem[{{Bullock} \& {Boylan-Kolchin}(2017)}]{2017ARA&A..55..343B}
{Bullock}, J.~S., \& {Boylan-Kolchin}, M. 2017, \araa, 55, 343

\bibitem[{{Bullock} {et~al.}(2000){Bullock}, {Kravtsov}, \& {Weinberg}}]{2000ApJ...539..517B}
{Bullock}, J.~S., {Kravtsov}, A.~V., \& {Weinberg}, D.~H. 2000, \apj, 539, 517

\bibitem[{{Choi} {et~al.}(2022){Choi}, {Ostriker}, {Hirschmann}, {Somerville}, \& {Naab}}]{2022ApJ...929..113C}
{Choi}, E., {Ostriker}, J.~P., {Hirschmann}, M., {Somerville}, R.~S., \& {Naab}, T. 2022, \apj, 929, 113

\bibitem[{Choi {et~al.}(2012)Choi, Ostriker, Naab, \& Johansson}]{Choi2012a}
Choi, E., Ostriker, J.~P., Naab, T., \& Johansson, P.~H. 2012, Astrophysical Journal, 754, 125

\bibitem[{Choi {et~al.}(2017)Choi, Ostriker, Naab, Somerville, Hirschmann, Núñez, Hu, \& Oser}]{Choi2017}
Choi, E., Ostriker, J.~P., Naab, T., {et~al.} 2017, The Astrophysical Journal, 844, 31

\bibitem[{Choi {et~al.}(2018)Choi, Somerville, Ostriker, Naab, \& Hirschmann}]{Choi2018}
Choi, E., Somerville, R.~S., Ostriker, J.~P., Naab, T., \& Hirschmann, M. 2018, The Astrophysical Journal, 866, 91

\bibitem[{{Cooper} {et~al.}(2010){Cooper}, {Cole}, {Frenk}, {White}, {Helly}, {Benson}, {De Lucia}, {Helmi}, {Jenkins}, {Navarro}, {Springel}, \& {Wang}}]{2010MNRAS.406..744C}
{Cooper}, A.~P., {Cole}, S., {Frenk}, C.~S., {et~al.} 2010, \mnras, 406, 744

\bibitem[{Costa {et~al.}(2019)Costa, Rosdahl, \& Kimm}]{Costa2019}
Costa, T., Rosdahl, J., \& Kimm, T. 2019, Monthly Notices of the Royal Astronomical Society, 489, 5181–5186

\bibitem[{Cullen \& Dehnen(2010)}]{2010MNRAS.408..669C}
Cullen, L., \& Dehnen, W. 2010, Monthly Notices of the Royal Astronomical Society, 408, 669

\bibitem[{Dashyan {et~al.}(2019)Dashyan, Choi, Somerville, Naab, Quirk, Hirschmann, \& Ostriker}]{Dashyan2019}
Dashyan, G., Choi, E., Somerville, R.~S., {et~al.} 2019, Monthly Notices of the Royal Astronomical Society, Volume 487, Issue 4, p.5889-5901, 487, 5889

\bibitem[{Dehnen \& Aly(2012)}]{2012MNRAS.425.1068D}
Dehnen, W., \& Aly, H. 2012, Monthly Notices of the Royal Astronomical Society, 425, 1068

\bibitem[{{Dubois} {et~al.}(2016){Dubois}, {Peirani}, {Pichon}, {Devriendt}, {Gavazzi}, {Welker}, \& {Volonteri}}]{2016MNRAS.463.3948D}
{Dubois}, Y., {Peirani}, S., {Pichon}, C., {et~al.} 2016, \mnras, 463, 3948

\bibitem[{Durier \& Vecchia(2012)}]{2012MNRAS.419..465D}
Durier, F., \& Vecchia, C.~D. 2012, Monthly Notices of the Royal Astronomical Society, 419, 465

\bibitem[{{Emerick} {et~al.}(2019){Emerick}, {Bryan}, \& {Mac Low}}]{2019MNRAS.482.1304E}
{Emerick}, A., {Bryan}, G.~L., \& {Mac Low}, M.-M. 2019, \mnras, 482, 1304

\bibitem[{{Font} {et~al.}(2022){Font}, {McCarthy}, {Belokurov}, {Brown}, \& {Stafford}}]{2022MNRAS.511.1544F}
{Font}, A.~S., {McCarthy}, I.~G., {Belokurov}, V., {Brown}, S.~T., \& {Stafford}, S.~G. 2022, \mnras, 511, 1544

\bibitem[{Gallazzi {et~al.}(2005)Gallazzi, Charlot, Brinchmann, White, \& Tremonti}]{Gallazzi2005}
Gallazzi, A., Charlot, S., Brinchmann, J., White, S. D.~M., \& Tremonti, C.~A. 2005, Monthly Notices of the Royal Astronomical Society, 362, 41–58

\bibitem[{{Geen} {et~al.}(2013){Geen}, {Slyz}, \& {Devriendt}}]{2013MNRAS.429..633G}
{Geen}, S., {Slyz}, A., \& {Devriendt}, J. 2013, \mnras, 429, 633

\bibitem[{{Geha} {et~al.}(2017){Geha}, {Wechsler}, {Mao}, {Tollerud}, {Weiner}, {Bernstein}, {Hoyle}, {Marchi}, {Marshall}, {Mu{\~n}oz}, \& {Lu}}]{Geha2017}
{Geha}, M., {Wechsler}, R.~H., {Mao}, Y.-Y., {et~al.} 2017, \apj, 847, 4

\bibitem[{{Geha} {et~al.}(2024){Geha}, {Mao}, {Wechsler}, {Asali}, {Kado-Fong}, {Kallivayalil}, {Nadler}, {Tollerud}, {Weiner}, {de los Reyes}, {Wang}, \& {Wu}}]{Geha2024}
{Geha}, M., {Mao}, Y.-Y., {Wechsler}, R.~H., {et~al.} 2024, \apj, 976, 118

\bibitem[{{Haardt} \& {Madau}(2001)}]{2001cghr.confE..64H}
{Haardt}, F., \& {Madau}, P. 2001, in Clusters of Galaxies and the High Redshift Universe Observed in X-rays, ed. D.~M. {Neumann} \& J.~T.~V. {Tran}, 64

\bibitem[{{Hambrick} {et~al.}(2011){Hambrick}, {Ostriker}, {Naab}, \& {Johansson}}]{2011ApJ...738...16H}
{Hambrick}, D.~C., {Ostriker}, J.~P., {Naab}, T., \& {Johansson}, P.~H. 2011, \apj, 738, 16

\bibitem[{{Hilz} {et~al.}(2013){Hilz}, {Naab}, \& {Ostriker}}]{2013MNRAS.429.2924H}
{Hilz}, M., {Naab}, T., \& {Ostriker}, J.~P. 2013, \mnras, 429, 2924

\bibitem[{{Hirschmann} {et~al.}(2015){Hirschmann}, {Naab}, {Ostriker}, {Forbes}, {Duc}, {Dav{\'e}}, {Oser}, \& {Karabal}}]{2015MNRAS.449..528H}
{Hirschmann}, M., {Naab}, T., {Ostriker}, J.~P., {et~al.} 2015, \mnras, 449, 528

\bibitem[{Hopkins(2013)}]{2013MNRAS.428.2840H}
Hopkins, P.~F. 2013, Monthly Notices of the Royal Astronomical Society, 428, 2840

\bibitem[{{Hopkins} {et~al.}(2010){Hopkins}, {Bundy}, {Hernquist}, {Wuyts}, \& {Cox}}]{2010MNRAS.401.1099H}
{Hopkins}, P.~F., {Bundy}, K., {Hernquist}, L., {Wuyts}, S., \& {Cox}, T.~J. 2010, \mnras, 401, 1099

\bibitem[{{Hopkins} {et~al.}(2014){Hopkins}, {Kere{\v{s}}}, {O{\~n}orbe}, {Faucher-Gigu{\`e}re}, {Quataert}, {Murray}, \& {Bullock}}]{2014MNRAS.445..581H}
{Hopkins}, P.~F., {Kere{\v{s}}}, D., {O{\~n}orbe}, J., {et~al.} 2014, \mnras, 445, 581

\bibitem[{{Hopkins} {et~al.}(2018){Hopkins}, {Wetzel}, {Kere{\v{s}}}, {Faucher-Gigu{\`e}re}, {Quataert}, {Boylan-Kolchin}, {Murray}, {Hayward}, \& {El-Badry}}]{2018MNRAS.477.1578H}
{Hopkins}, P.~F., {Wetzel}, A., {Kere{\v{s}}}, D., {et~al.} 2018, \mnras, 477, 1578

\bibitem[{Hoyle \& Lyttleton(1939)}]{1939PCPS...34..405H}
Hoyle, F., \& Lyttleton, R.~A. 1939, Proceedings of the Cambridge Philosophical Society, 34, 405

\bibitem[{Hu {et~al.}(2014)Hu, Naab, Walch, Moster, \& Oser}]{2014MNRAS.443.1173H}
Hu, C.~Y., Naab, T., Walch, S., Moster, B.~P., \& Oser, L. 2014, Monthly Notices of the Royal Astronomical Society, 443, 1173

\bibitem[{{Huang} {et~al.}(2013){Huang}, {Ho}, {Peng}, {Li}, \& {Barth}}]{2013ApJ...768L..28H}
{Huang}, S., {Ho}, L.~C., {Peng}, C.~Y., {Li}, Z.-Y., \& {Barth}, A.~J. 2013, \apjl, 768, L28

\bibitem[{Hui {et~al.}(2022)Hui, Peng, Ying-zhong, Zhi-quan, Han, \& Cai-ping}]{Hui2022}
Hui, G.~U., Peng, W.~A., Ying-zhong, X.~U., {et~al.} 2022, Chinese Astronomy and Astrophysics, 46, doi:10.1016/j.chinastron.2022.11.003

\bibitem[{{Ibrahim} \& {Kobayashi}(2024)}]{2024MNRAS.527.3276I}
{Ibrahim}, D., \& {Kobayashi}, C. 2024, \mnras, 527, 3276

\bibitem[{Iwamoto {et~al.}(1999)Iwamoto, Brachwitz, Nomoto, Kishimoto, Umeda, Hix, \& Thielemann}]{1999ApJS..125..439I}
Iwamoto, K., Brachwitz, F., Nomoto, K., {et~al.} 1999, The Astrophysical Journal Supplement Series, 125, 439

\bibitem[{Karakas(2010)}]{2010MNRAS.403.1413K}
Karakas, A.~I. 2010, Monthly Notices of the Royal Astronomical Society, 403, 1413

\bibitem[{{Katz} {et~al.}(2020){Katz}, {Ramsoy}, {Rosdahl}, {Kimm}, {Blaizot}, {Haehnelt}, {Michel-Dansac}, {Garel}, {Laigle}, {Devriendt}, \& {Slyz}}]{2020MNRAS.494.2200K}
{Katz}, H., {Ramsoy}, M., {Rosdahl}, J., {et~al.} 2020, \mnras, 494, 2200

\bibitem[{{Kauffmann} \& {Haehnelt}(2000)}]{2000MNRAS.311..576K}
{Kauffmann}, G., \& {Haehnelt}, M. 2000, \mnras, 311, 576

\bibitem[{{Kim} \& {Ostriker}(2015)}]{2015ApJ...802...99K}
{Kim}, C.-G., \& {Ostriker}, E.~C. 2015, \apj, 802, 99

\bibitem[{{Kimm} {et~al.}(2015){Kimm}, {Cen}, {Devriendt}, {Dubois}, \& {Slyz}}]{2015MNRAS.451.2900K}
{Kimm}, T., {Cen}, R., {Devriendt}, J., {Dubois}, Y., \& {Slyz}, A. 2015, \mnras, 451, 2900

\bibitem[{{Klypin} {et~al.}(1999){Klypin}, {Kravtsov}, {Valenzuela}, \& {Prada}}]{1999ApJ...522...82K}
{Klypin}, A., {Kravtsov}, A.~V., {Valenzuela}, O., \& {Prada}, F. 1999, \apj, 522, 82

\bibitem[{{Koposov} {et~al.}(2008){Koposov}, {Belokurov}, {Evans}, {Hewett}, {Irwin}, {Gilmore}, {Zucker}, {Rix}, {Fellhauer}, {Bell}, \& {Glushkova}}]{2008ApJ...686..279K}
{Koposov}, S., {Belokurov}, V., {Evans}, N.~W., {et~al.} 2008, \apj, 686, 279

\bibitem[{{Koposov} {et~al.}(2015){Koposov}, {Belokurov}, {Torrealba}, \& {Evans}}]{2015ApJ...805..130K}
{Koposov}, S.~E., {Belokurov}, V., {Torrealba}, G., \& {Evans}, N.~W. 2015, \apj, 805, 130

\bibitem[{{Kravtsov} {et~al.}(2004){Kravtsov}, {Gnedin}, \& {Klypin}}]{2004ApJ...609..482K}
{Kravtsov}, A.~V., {Gnedin}, O.~Y., \& {Klypin}, A.~A. 2004, \apj, 609, 482

\bibitem[{{Kravtsov} {et~al.}(2018){Kravtsov}, {Vikhlinin}, \& {Meshcheryakov}}]{Kravtsov2018}
{Kravtsov}, A.~V., {Vikhlinin}, A.~A., \& {Meshcheryakov}, A.~V. 2018, Astronomy Letters, 44, 8

\bibitem[{{Macci{\`o}} {et~al.}(2010){Macci{\`o}}, {Kang}, {Fontanot}, {Somerville}, {Koposov}, \& {Monaco}}]{2010MNRAS.402.1995M}
{Macci{\`o}}, A.~V., {Kang}, X., {Fontanot}, F., {et~al.} 2010, \mnras, 402, 1995

\bibitem[{{Makarov} \& {Karachentsev}(2011)}]{2011MNRAS.412.2498M}
{Makarov}, D., \& {Karachentsev}, I. 2011, \mnras, 412, 2498

\bibitem[{{Mao} {et~al.}(2024){Mao}, {Geha}, {Wechsler}, {Asali}, {Wang}, {Kado-Fong}, {Kallivayalil}, {Nadler}, {Tollerud}, {Weiner}, {de los Reyes}, \& {Wu}}]{Mao2024}
{Mao}, Y.-Y., {Geha}, M., {Wechsler}, R.~H., {et~al.} 2024, \apj, 976, 117

\bibitem[{{M{\'a}rmol-Queralt{\'o}} {et~al.}(2012){M{\'a}rmol-Queralt{\'o}}, {Trujillo}, {P{\'e}rez-Gonz{\'a}lez}, {Varela}, \& {Barro}}]{2012MNRAS.422.2187M}
{M{\'a}rmol-Queralt{\'o}}, E., {Trujillo}, I., {P{\'e}rez-Gonz{\'a}lez}, P.~G., {Varela}, J., \& {Barro}, G. 2012, \mnras, 422, 2187

\bibitem[{{McConnachie}(2012)}]{2012AJ....144....4M}
{McConnachie}, A.~W. 2012, \aj, 144, 4

\bibitem[{{McConnachie} \& {Irwin}(2006)}]{2006MNRAS.365.1263M}
{McConnachie}, A.~W., \& {Irwin}, M.~J. 2006, \mnras, 365, 1263

\bibitem[{{Moore} {et~al.}(1999){Moore}, {Ghigna}, {Governato}, {Lake}, {Quinn}, {Stadel}, \& {Tozzi}}]{1999ApJ...524L..19M}
{Moore}, B., {Ghigna}, S., {Governato}, F., {et~al.} 1999, \apjl, 524, L19

\bibitem[{Moster {et~al.}(2013)Moster, Naab, \& White}]{2013MNRAS.428.3121M}
Moster, B.~P., Naab, T., \& White, S.~D.~M. 2013, \mnras, 428, 3121

\bibitem[{Mármol-Queraltó {et~al.}(2013)Mármol-Queraltó, Trujillo, Villar, Barro, \& Pérez-González}]{2013MNRAS.429..792M}
Mármol-Queraltó, E., Trujillo, I., Villar, V., Barro, G., \& Pérez-González, P.~G. 2013, \mnras, 429, 792

\bibitem[{{Naab} {et~al.}(2009){Naab}, {Johansson}, \& {Ostriker}}]{2009ApJ...699L.178N}
{Naab}, T., {Johansson}, P.~H., \& {Ostriker}, J.~P. 2009, \apjl, 699, L178

\bibitem[{{Naab} {et~al.}(2007){Naab}, {Johansson}, {Ostriker}, \& {Efstathiou}}]{2007ApJ...658..710N}
{Naab}, T., {Johansson}, P.~H., {Ostriker}, J.~P., \& {Efstathiou}, G. 2007, \apj, 658, 710

\bibitem[{{Naab} \& {Ostriker}(2017)}]{2017ARA&A..55...59N}
{Naab}, T., \& {Ostriker}, J.~P. 2017, \araa, 55, 59

\bibitem[{{Nierenberg} {et~al.}(2012){Nierenberg}, {Auger}, {Treu}, {Marshall}, {Fassnacht}, \& {Busha}}]{2012ApJ...752...99N}
{Nierenberg}, A.~M., {Auger}, M.~W., {Treu}, T., {et~al.} 2012, \apj, 752, 99

\bibitem[{{Nierenberg} {et~al.}(2013){Nierenberg}, {Treu}, {Menci}, {Lu}, \& {Wang}}]{2013ApJ...772..146N}
{Nierenberg}, A.~M., {Treu}, T., {Menci}, N., {Lu}, Y., \& {Wang}, W. 2013, \apj, 772, 146

\bibitem[{{Nipoti} {et~al.}(2009){Nipoti}, {Treu}, {Auger}, \& {Bolton}}]{Nipoti2009}
{Nipoti}, C., {Treu}, T., {Auger}, M.~W., \& {Bolton}, A.~S. 2009, \apjl, 706, L86

\bibitem[{Nyman {et~al.}(1992)Nyman, Booth, Carlström, Habing, Heske, Sahai, Stark, van~der Veen, \& Winnberg}]{1992A&amp;AS...93..121N}
Nyman, L., Booth, R., Carlström, U., {et~al.} 1992, Astronomy and Astrophysics Supplement Series, 93, 121

\bibitem[{Núñez {et~al.}(2017)Núñez, Ostriker, Naab, Oser, Hu, \& Choi}]{2017ApJ...836..204N}
Núñez, A., Ostriker, J.~P., Naab, T., {et~al.} 2017, The Astrophysical Journal, 836, 204

\bibitem[{{Okamoto} \& {Frenk}(2009)}]{2009MNRAS.399L.174O}
{Okamoto}, T., \& {Frenk}, C.~S. 2009, \mnras, 399, L174

\bibitem[{{Okamoto} {et~al.}(2010){Okamoto}, {Frenk}, {Jenkins}, \& {Theuns}}]{2010MNRAS.406..208O}
{Okamoto}, T., {Frenk}, C.~S., {Jenkins}, A., \& {Theuns}, T. 2010, \mnras, 406, 208

\bibitem[{{Okamoto} {et~al.}(2008){Okamoto}, {Gao}, \& {Theuns}}]{2008MNRAS.390..920O}
{Okamoto}, T., {Gao}, L., \& {Theuns}, T. 2008, \mnras, 390, 920

\bibitem[{Oser {et~al.}(2010)Oser, Ostriker, Naab, Johansson, \& Burkert}]{Oser2010}
Oser, L., Ostriker, J.~P., Naab, T., Johansson, P.~H., \& Burkert, A. 2010, Astrophysical Journal, 725, 2312

\bibitem[{Panter {et~al.}(2008)Panter, Jimenez, Heavens, \& Charlot}]{Panter2008}
Panter, B., Jimenez, R., Heavens, A.~F., \& Charlot, S. 2008, Monthly Notices of the Royal Astronomical Society, 391, 1117–1126

\bibitem[{Pasquali {et~al.}(2010)Pasquali, Gallazzi, Fontanot, Van Den~Bosch, De~Lucia, Mo, \& Yang}]{Pasquali2010}
Pasquali, A., Gallazzi, A., Fontanot, F., {et~al.} 2010, Monthly Notices of the Royal Astronomical Society, 407, 937–954

\bibitem[{Quilis \& Trujillo(2012)}]{2012ApJ...752L..19Q}
Quilis, V., \& Trujillo, I. 2012, The Astrophysical Journal Letters, 752, L19

\bibitem[{Read \& Hayfield(2012)}]{2012MNRAS.422.3037R}
Read, J.~I., \& Hayfield, T. 2012, Monthly Notices of the Royal Astronomical Society, 422, 3037

\bibitem[{Ritchie \& Thomas(2001)}]{2001MNRAS.323..743R}
Ritchie, B.~W., \& Thomas, P.~A. 2001, Monthly Notices of the Royal Astronomical Society, 323, 743

\bibitem[{Rohr {et~al.}(2024)Rohr, Pillepich, Nelson, Ayromlou, \& Zinger}]{Rohr2024}
Rohr, E., Pillepich, A., Nelson, D., Ayromlou, M., \& Zinger, E. 2024, Astronomy and Astrophysics, 686, A86

\bibitem[{Ruiz {et~al.}(2014)Ruiz, Trujillo, \& Mármol-Queraltó}]{2014MNRAS.442..347R}
Ruiz, P., Trujillo, I., \& Mármol-Queraltó, E. 2014, \mnras, 442, 347

\bibitem[{Ruiz {et~al.}(2015)Ruiz, Trujillo, \& Mármol-Queraltó}]{2015MNRAS.454.1605R}
---. 2015, \mnras, 454, 1605

\bibitem[{Röttgers {et~al.}(2020)Röttgers, Naab, Cernetic, Davé, Kauffmann, Borthakur, \& Foidl}]{Rottgers2020a}
Röttgers, B., Naab, T., Cernetic, M., {et~al.} 2020, Monthly Notices of the Royal Astronomical Society, 496, 152

\bibitem[{Saitoh \& Makino(2009)}]{2009ApJ...697L..99S}
Saitoh, T.~R., \& Makino, J. 2009, Astrophysical Journal, 697, L99

\bibitem[{Saitoh \& Makino(2013)}]{2013ApJ...768...44S}
---. 2013, Astrophysical Journal, 768, 44

\bibitem[{{Samuel} {et~al.}(2022){Samuel}, {Wetzel}, {Santistevan}, {Tollerud}, {Moreno}, {Boylan-Kolchin}, {Bailin}, \& {Pardasani}}]{2022MNRAS.514.5276S}
{Samuel}, J., {Wetzel}, A., {Santistevan}, I., {et~al.} 2022, \mnras, 514, 5276

\bibitem[{{Samuel} {et~al.}(2020){Samuel}, {Wetzel}, {Tollerud}, {Garrison-Kimmel}, {Loebman}, {El-Badry}, {Hopkins}, {Boylan-Kolchin}, {Faucher-Gigu{\`e}re}, {Bullock}, {Benincasa}, \& {Bailin}}]{2020MNRAS.491.1471S}
{Samuel}, J., {Wetzel}, A., {Tollerud}, E., {et~al.} 2020, \mnras, 491, 1471

\bibitem[{{Sawala} {et~al.}(2016){Sawala}, {Frenk}, {Fattahi}, {Navarro}, {Bower}, {Crain}, {Dalla Vecchia}, {Furlong}, {Helly}, {Jenkins}, {Oman}, {Schaller}, {Schaye}, {Theuns}, {Trayford}, \& {White}}]{2016MNRAS.457.1931S}
{Sawala}, T., {Frenk}, C.~S., {Fattahi}, A., {et~al.} 2016, \mnras, 457, 1931

\bibitem[{Sazonov {et~al.}(2005)Sazonov, Ostriker, Ciotti, \& Sunyaev}]{2005MNRAS.358..168S}
Sazonov, S.~Y., Ostriker, J.~P., Ciotti, L., \& Sunyaev, R.~A. 2005, Monthly Notices of the Royal Astronomical Society, 358, 168

\bibitem[{Sazonov {et~al.}(2004)Sazonov, Ostriker, \& Sunyaev}]{2004MNRAS.347..144S}
Sazonov, S.~Y., Ostriker, J.~P., \& Sunyaev, R.~A. 2004, Monthly Notices of the Royal Astronomical Society, 347, 144

\bibitem[{{Scannapieco} {et~al.}(2008){Scannapieco}, {Tissera}, {White}, \& {Springel}}]{2008MNRAS.389.1137S}
{Scannapieco}, C., {Tissera}, P.~B., {White}, S. D.~M., \& {Springel}, V. 2008, \mnras, 389, 1137

\bibitem[{{Shen} {et~al.}(2014){Shen}, {Madau}, {Conroy}, {Governato}, \& {Mayer}}]{2014ApJ...792...99S}
{Shen}, S., {Madau}, P., {Conroy}, C., {Governato}, F., \& {Mayer}, L. 2014, \apj, 792, 99

\bibitem[{{Simpson} {et~al.}(2015){Simpson}, {Bryan}, {Hummels}, \& {Ostriker}}]{2015ApJ...809...69S}
{Simpson}, C.~M., {Bryan}, G.~L., {Hummels}, C., \& {Ostriker}, J.~P. 2015, \apj, 809, 69

\bibitem[{{Simpson} {et~al.}(2018){Simpson}, {Grand}, {G{\'o}mez}, {Marinacci}, {Pakmor}, {Springel}, {Campbell}, \& {Frenk}}]{2018MNRAS.478..548S}
{Simpson}, C.~M., {Grand}, R. J.~J., {G{\'o}mez}, F.~A., {et~al.} 2018, \mnras, 478, 548

\bibitem[{{Somerville} \& {Dav{\'e}}(2015)}]{2015ARA&A..53...51S}
{Somerville}, R.~S., \& {Dav{\'e}}, R. 2015, \araa, 53, 51

\bibitem[{{Somerville} {et~al.}(2008){Somerville}, {Hopkins}, {Cox}, {Robertson}, \& {Hernquist}}]{2008MNRAS.391..481S}
{Somerville}, R.~S., {Hopkins}, P.~F., {Cox}, T.~J., {Robertson}, B.~E., \& {Hernquist}, L. 2008, \mnras, 391, 481

\bibitem[{Spergel {et~al.}(2007)Spergel, Bean, Dore, Nolta, Bennett, Dunkley, Hinshaw, Jarosik, Komatsu, Page, Peiris, Verde, Halpern, Hill, Kogut, Limon, Meyer, Odegard, Tucker, Weiland, Wollack, \& Wright}]{2007ApJS..170..377S}
Spergel, D.~N., Bean, R., Dore, O., {et~al.} 2007, The Astrophysical Journal Supplement Series, 170, 377

\bibitem[{Springel(2005)}]{2005MNRAS.364.1105S}
Springel, V. 2005, Monthly Notices of the Royal Astronomical Society, 364, 1105

\bibitem[{{Springel} {et~al.}(2008){Springel}, {Wang}, {Vogelsberger}, {Ludlow}, {Jenkins}, {Helmi}, {Navarro}, {Frenk}, \& {White}}]{2008MNRAS.391.1685S}
{Springel}, V., {Wang}, J., {Vogelsberger}, M., {et~al.} 2008, \mnras, 391, 1685

\bibitem[{Strömgren(1939)}]{1939ApJ....89..526S}
Strömgren, B. 1939, The Astrophysical Journal, 89, 526

\bibitem[{Tomozeiu {et~al.}(2016{\natexlab{a}})Tomozeiu, Mayer, \& Quinn}]{2016ApJ...818..193T}
Tomozeiu, M., Mayer, L., \& Quinn, T. 2016{\natexlab{a}}, The Astrophysical Journal, 818, 193

\bibitem[{Tomozeiu {et~al.}(2016{\natexlab{b}})Tomozeiu, Mayer, \& Quinn}]{2016ApJ...827L..15T}
---. 2016{\natexlab{b}}, The Astrophysical Journal Letters, 827, L15

\bibitem[{Vogelsberger {et~al.}(2014)Vogelsberger, Genel, Springel, Torrey, Sijacki, Xu, Snyder, Bird, Nelson, \& Hernquist}]{2014Natur.509..177V}
Vogelsberger, M., Genel, S., Springel, V., {et~al.} 2014, Nature, 509, 177

\bibitem[{{Walch} \& {Naab}(2015)}]{2015MNRAS.451.2757W}
{Walch}, S., \& {Naab}, T. 2015, \mnras, 451, 2757

\bibitem[{{Wang} {et~al.}(2024){Wang}, {Nadler}, {Mao}, {Wechsler}, {Abel}, {Behroozi}, {Geha}, {Asali}, {de los Reyes}, {Kado-Fong}, {Kallivayalil}, {Tollerud}, {Weiner}, \& {Wu}}]{2024arXiv240414500W}
{Wang}, Y., {Nadler}, E.~O., {Mao}, Y.-Y., {et~al.} 2024, \apj, 976, 119

\bibitem[{{Wetzel} {et~al.}(2016){Wetzel}, {Hopkins}, {Kim}, {Faucher-Gigu{\`e}re}, {Kere{\v{s}}}, \& {Quataert}}]{2016ApJ...827L..23W}
{Wetzel}, A.~R., {Hopkins}, P.~F., {Kim}, J.-h., {et~al.} 2016, \apjl, 827, L23

\bibitem[{{Wetzel} {et~al.}(2015){Wetzel}, {Tollerud}, \& {Weisz}}]{2015ApJ...808L..27W}
{Wetzel}, A.~R., {Tollerud}, E.~J., \& {Weisz}, D.~R. 2015, \apjl, 808, L27

\bibitem[{Wiersma {et~al.}(2009)Wiersma, Schaye, \& Smith}]{2009MNRAS.393...99W}
Wiersma, R.~P., Schaye, J., \& Smith, B.~D. 2009, Monthly Notices of the Royal Astronomical Society, 393, 99

\bibitem[{Woosley \& Weaver(1995)}]{1995ApJS..101..181W}
Woosley, S.~E., \& Weaver, T.~A. 1995, The Astrophysical Journal Supplement Series, 101, 181

\bibitem[{Wu {et~al.}(2022)Wu, Peek, Tollerud, Mao, Nadler, Geha, Wechsler, Kallivayalil, \& Weiner}]{Wu2022}
Wu, J.~F., Peek, J. E.~G., Tollerud, E.~J., {et~al.} 2022, The Astrophysical Journal, 927, 121

\bibitem[{{Zhu} {et~al.}(2024){Zhu}, {Tonnesen}, \& {Bryan}}]{2024ApJ...960...54Z}
{Zhu}, J., {Tonnesen}, S., \& {Bryan}, G.~L. 2024, \apj, 960, 54

\bibitem[{{Zolotov} {et~al.}(2012){Zolotov}, {Brooks}, {Willman}, {Governato}, {Pontzen}, {Christensen}, {Dekel}, {Quinn}, {Shen}, \& {Wadsley}}]{2012ApJ...761...71Z}
{Zolotov}, A., {Brooks}, A.~M., {Willman}, B., {et~al.} 2012, \apj, 761, 71

\end{thebibliography}

\end{document}